\documentclass[10pt,twocolumn,aps,prb,amsmath,amssymb,superscriptaddress,floatfix]{revtex4-2}

\usepackage{graphicx}
\usepackage{color}
\usepackage{txfonts}
\usepackage[
colorlinks,
citecolor=blue,
linkcolor=blue,
urlcolor=blue,plainpages=false,pdfpagelabels
]{hyperref}

\newcommand{\be}{\begin{equation}}
\newcommand{\ee}{\end{equation}}
\newcommand{\bea}{\begin{eqnarray}}
\newcommand{\eea}{\end{eqnarray}}
\newcommand{\lb}{\left[}
\newcommand{\rb}{\right]}
\newcommand{\lp}{\left(}
\newcommand{\rp}{\right)}
\newcommand{\eps}{\varepsilon}
\newcommand{\zero}{{(0)}}
\newcommand{\integral}[3]{\int_{#1}^{#2} #3 \hspace{1pt}}

\renewcommand{\tilde}{\widetilde}
\renewcommand{\vec}[1]{{\bf #1}}

\def\nn{\nonumber\\}

\graphicspath{{images/}}

\begin{document}

\title{Electron hydrodynamics of anomalous Hall materials }

\author{Eddwi H. Hasdeo}
\email{hesky.hasdeo@uni.lu}
\affiliation{Department of Physics and Material Science, University of Luxembourg, Luxembourg}
\affiliation{Research Center for Physics, Indonesian Institute of Sciences, South Tangerang, Indonesia}

\author{Johan Ekstr\"om}
\affiliation{Department of Physics and Material Science, University of Luxembourg, Luxembourg}

\author{Edvin G. Idrisov}
\affiliation{Department of Physics and Material Science, University of Luxembourg, Luxembourg}

\author{Thomas L. Schmidt}
\email{thomas.schmidt@uni.lu}
\affiliation{Department of Physics and Material Science, University of Luxembourg, Luxembourg}

\begin{abstract}
  We study two-dimensional electron systems in the hydrodynamic
  regime. We show that a geometrical Berry curvature modifies the
  effective Navier-Stokes equation for viscous electron flow in
  topological materials. For small electric fields, the Hall current
  becomes negligible compared to the viscous longitudinal current. In
  this regime, we highlight an unconventional Poiseuille flow with
  asymmetric profile and a deviation of the maximum of the current
  from the center of the system. In a two-dimensional infinite
  geometry, the Berry curvature leads to current whirlpools and an
  asymmetry of potential profile. This phenomenon can be probed by
  measuring the asymmetric non-local resistance profile.
\end{abstract}

\maketitle
\section{Introduction}
The collective motion of electrons in metals can lead to starkly different physical effects from those in the more conventional free-electron transport. One
particular example is electron hydrodynamics, where the flow of
electrons resembles that of a viscous fluid. This hydrodynamic regime can be reached at
intermediate temperatures ($T \sim 100\text{K}$) in ultraclean samples when the
rate of electron-electron scattering that conserves the electrons' energy
and momentum is larger than the momentum-relaxing scattering rates
due to phonons and impurities. In this regime,
the resistance $R$ decreases with increasing temperature $T$, in stark contrast to the proportionality of $R$ to $T$ in ordinary metals caused by electron-phonon
scattering~\cite{gurzhi1963,DeJong1995}. Moreover, when passing
through narrow constrictions, the conductance of viscous electron fluids can
exceed that in the ballistic limit of free electrons~\cite{Guo2017,Kumar2017}. Finally, the motion of viscous electrons can create non-local
potential disturbances leading to negative non-local resistance and current whirlpools due to electron backflow~\cite{Bandurin2016,Levitov2016,Pellegrino2016}.

Under certain approximations, a macroscopic hydrodynamic theory can be derived from the
``microscopic'' semiclassical Boltzmann transport equation~\cite{Lucas2018a,Narozhny2019}. By taking averages of
microscopic quantities over a local equilibrium distribution, one can
obtain the dynamics of the velocity field $\vec u(\vec r,t)$ which turns out to
obey the Euler equation or, when viscosity
is taken into account, the Navier-Stokes equation (NSE). For graphene-like systems, the structure of this NSE is interesting because it contains features from the relativistic Dirac-like spectrum of graphene. However, it mainly leads to hydrodynamic phenomena similar to those observed in classical, non-relativistic fluids, like for instance a Poiseuille flow profile,
even for systems that show many interesting quantum phenomena at the single-particle level like graphene.

Therefore, it is still an open question to what extent quantum mechanical features of the band structure manifest themselves in the semiclassical hydrodynamic transport regime. To elucidate this question, we study the hydrodynamic flow of electrons in graphene-like systems with a nonzero Berry curvature. The geometrical Berry curvature encodes the internal structure of the crystal wavefunctions and can, for instance, drive electrons perpendicular to
an applied electric field~\cite{xiao10}. This Berry curvature is the intrinsic cause of anomalous Hall transport in
multiband systems with either broken inversion or time-reversal
symmetry (TRS). We will show that in hydrodynamic metals with Berry curvature, the NSE will
reflect quantum effects due to the Berry curvature, with one of its consequences being a non-trivial
electron flow profile.

In this work, we explore the effects of Berry curvature on electron
hydrodynamics in systems with broken TRS. We note that our
results are distinct from the hydrodynamics in system with TRS studied in
Refs.~\onlinecite{Toshio2020,Tavakol2020} because the Berry curvature in systems with (without) TRS is an odd (even) function of momentum. We
derive the NSE including the Berry curvature and use it to study Poiseuille flow in a channel geometry as well as vortex formation in an experimentally relevant 2D half-plane geometry.

Concerning the Poiseuille flow, we find that the Berry curvature gives rise to an asymmetric velocity
profile where the maximum flow velocity deviates from the center of the channel. Moreover, although the ordinary NSE allows for non-zero vorticity, current whirlpools or vortices are generally absent in infinite 2D systems (half-plane geometry) in the absence of Berry curvature. In this case, current whirlpools arise from the possible current backflow in finite systems~\cite{Pellegrino2016}. In contrast, we show in this work that the Berry curvature alters
the vorticity equation and can produce sizable current whirlpools even
in an infinite geometry. These features are reflected directly in the non-local resistance, which can thus be used as an experimental tool to measure the influence of Berry curvature on the hydrodynamic flow.

\section{The Navier Stokes equation in topological materials}
We begin by considering a material with a Fermi level
in the conduction band and possessing a Berry curvature $\boldsymbol \Omega_\vec
p$. We allow TRS to be broken such that $\boldsymbol
\Omega_\vec p$ is an even function of the lattice momentum $\vec
p$. Under an externally applied electric field, the electron distribution function $f(\vec
p,\vec r, t)$ evolves according to the semiclassical Boltzmann equation,
\be
\frac {\partial f}{\partial t} + \dot{\vec r} \cdot \frac{\partial
  f}{\partial \vec r} +\dot{\vec p} \cdot \frac{\partial f}{\partial
  \vec p} = C[f], \label{eq:boltz}
\ee
where momentum and position evolve according to $\dot{\vec p} = e \vec E$ and $\dot{\vec r}=\vec v_g + \vec v_a$. Here,
$v_g=\partial_\vec p \eps_\vec p$ is the group velocity which is determined by the dispersion relation $\eps_{\vec{p}}$, $\vec v_a=
(e/\hbar)\vec E \times \vec \Omega_\vec p$ is the anomalous velocity due to the Berry curvature, and
$C$ is the collision integral due to scattering processes. In the hydrodynamic regime, $C$ is dominated by electron-electron interactions which conserve momentum. Scattering
from impurities and phonons are small perturbations and can be incorporated using the relaxation-time
approximation. Umklapp scattering is neglected because it requires a large momentum transfer. The strong interactions lead to a fast equilibration of the electrons and their local distribution function satisfies
\be
  f(\vec p, \vec r, t)=\frac{1}{\exp\left[\beta(\vec r, t)(\eps_\vec p-\mu(\vec r,t)-\vec u(\vec r,t)\cdot\vec p)\right] +1}, \label{eq:dist}
  \ee
which is similar to the Fermi-Dirac distribution with an additional drift velocity $\vec u$. Note that the inverse temperature $\beta$, as well as the chemical potential $\mu$ and the drift velocity $\vec u$ generally depend on position and time. The distribution function \eqref{eq:dist} ensures that the collision integral due to electron-electron interactions in Eq.~\eqref{eq:boltz} vanishes~\cite{Narozhny2019}. A finite relaxation time or relaxation length causes the true distribution function to slightly deviate from the local equilibrium distribution and results in dissipative corrections which give rise to a nonzero electron viscosity~\cite{Briskot2015,Principi2016}. In this work, we assume that the Berry curvature does not contribute significantly to the viscosity.

To derive hydrodynamic equations, we define the macroscopic number density, the energy density and the momentum density, respectively, as $\mathcal{O}(\vec r, t) = \sum_\vec p \chi(\vec p) f(\vec p, \vec r, t) $ where $\chi=\{1,\eps_\vec p, \vec p\}$ and $\mathcal{O}=\{N,N_\epsilon,\bar{\vec p}\}$. Moreover, we define the corresponding particle current density, energy current density, and the stress tensor, respectively, as $\mathcal{V}(\vec r, t)=\mathcal{V}^\zero+\mathcal{V}^a=\sum_\vec p [\vec v_g(\vec p) + \vec v_a(\vec p)] \chi(\vec p) f(\vec p, \vec r, t) $ where $\mathcal{V}=\{\vec J, \vec J_\eps,  \Pi\}$. We can then employ Eq.~\eqref{eq:boltz} to obtain continuity equations,
 \bea
  \partial_t N +\nabla \cdot \vec J& =&0,\nn
  \partial_t N_\eps+\nabla \cdot \vec J_\eps   &=& e\vec E \cdot \vec J^{(0)}, \nn
  \partial_t \bar{p}_i + \nabla_j   \Pi_{i,j} &=&  eN E_i.\label{eq:pre-euler}
  \eea
The non-zero values on the right-hand side of Eq.~\eqref{eq:pre-euler} imply that the momentum and energy flows are not conserved due to the applied external force and Joule heating, respectively.

Our next goal is to describe the dynamics of the velocity field $\vec u$ by relating it to the macroscopic quantities in Eq.~\eqref{eq:pre-euler}. Using the local equilibrium distribution~\eqref{eq:dist}, we obtain the following relations
\bea
\vec J^{(0)} &=& N \vec u\label{eq:cont},\\
\vec J_{\eps}^{(0)} &=& W \vec u,\quad W = N_\eps + P \label{eq:enthalpy}
\eea
where the subscripts $(0)$ indicate that these functions are defined for the local equilibrium distribution function. Here, $W$ is the enthalpy density and $P= k_B T\sum_\vec p {\rm ln} \left[ 1+ e^{-\beta(\eps_\vec p - \vec u \cdot \vec p -\mu)}\right]$ is the pressure. We note that Eqs.~\eqref{eq:cont} and \eqref{eq:enthalpy} are very general and are valid for an arbitrary dispersion relation $\eps_\vec p$ of the system.

\begin{figure*}
    \centering
    \includegraphics{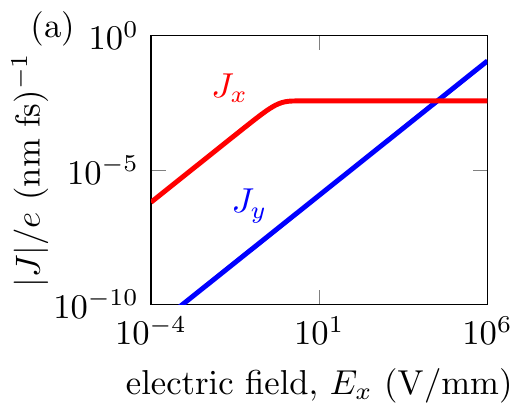}
  \hspace{0.1cm}
  \includegraphics{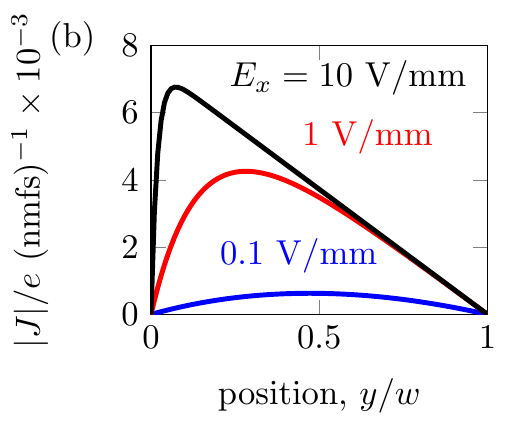}
  \hspace{0.1cm}
  \includegraphics{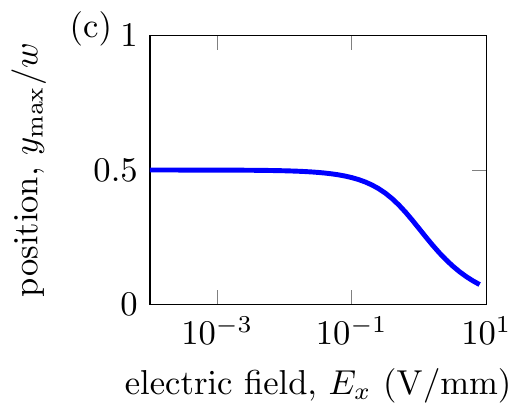}
\caption{\emph{Current profile in viscous Hall materials}. (a) Longitudinal current at $y=w/2$ (red) and Hall  current (blue) as a function of electric field. (b) Longitudinal current profile as a function of position for several values of electric field. (c) Maximum current position $y_{max}$ as a function of electric field. In (a)--(c) we have used a gapped Dirac model with $\Delta=\rm 100\ meV$, $T=100\ \rm K$, $w=1\rm \ \mu m$ and $\eta=6\times 10^{-3}\hbar N$ and $\mu$ is at the bottom of the conduction band. \label{Fig1}}
\end{figure*}

In contrast, the momentum equation in Eq.~\eqref{eq:pre-euler} depends explicitly on the structure of $\eps_\vec p$. Therefore, we will focus on an anomalous Hall system described by a gapped 2D Dirac Hamiltonian $H = \vec d \cdot \boldsymbol{\tau}$, with $\boldsymbol{\tau}$ being
the vector of Pauli matrices and $\vec d = (vp_x,vp_y, \Delta)$. Here, $v$ is the Fermi velocity and $2\Delta$ is the band gap. This system has
the dispersion relation $\eps^\pm_\vec p=\pm \sqrt{v^2p^2+\Delta^2}$ and the
Berry curvature $\boldsymbol\Omega_\vec p^\pm = v^2\hbar^2 \Delta/\lb 2(\eps_\vec p^\pm)^3\rb\hat{\vec z}$. A
single Dirac cone reflects the broken TRS that
can be found in magnetically doped topological insulators such as $\rm Cr$--doped bismuth telluride~\cite{Chang2013} or honeycomb crystals
(graphene, silicene, transition metal dichalcogenides, etc.) that sit
on top of a magnetic substrate~\cite{Wang2015,Tang2018,Averyanov2018}. Using the specific local distribution~\eqref{eq:dist} and the Dirac energy dispersion $\eps_\vec p$, we find
\be
\bar{\vec p} = v^{-2} W \vec u,\label{eq:np}\quad
  \Pi_{i,j}^{(0)}= P \delta_{i,j} + \frac{W}{v^2} u_i u_j.
\ee
The forms of these observables coincide with the corresponding expressions for a gapless Dirac dispersion in graphene~\cite{Narozhny2019}.

The anomalous quantities in Eq.~\eqref{eq:pre-euler} do not have a simple relationship with $\vec u$. Therefore, to make progress, we write the distribution function $f$ in Eq.~\eqref{eq:dist} as,
\be
f = f^{0}+\delta f,\quad  \delta f = \frac{\partial f^0}{\partial \eps} \lp \frac{\epsilon-\mu}{\beta}\delta \beta -\delta \mu - \vec p \cdot \vec u\rp, \label{eq:perturbdist}
\ee
where $f^{0}$ is the equilibrium Fermi-Dirac distribution function, corresponding to constant $\beta$ and $\mu$, as well as $\vec u = 0$. Moreover, $\delta f$ is a small perturbation accounting for the nonequilibrium state of the system. If we limit ourselves to the dynamics of linear order in $\vec u$, we can assume $\delta \beta$ to be small and $\delta \mu$ can be absorbed into the external electric field. In Eq.~\eqref{eq:perturbdist}, the terms containing $\delta \beta$ and $\delta \mu$ are even in $\vec p$ while the term containing $\vec u$ is odd in $\vec p$.

We note here that within the linear-response regime, the anomalous velocity does not give rise to a contribution to the continuity equations for the particle and energy current. Indeed, defining the electric potential $\phi$ via $\vec E=-\nabla \phi$, one finds that $\nabla\cdot \vec J^a=0$ and $\nabla\cdot \vec J^a_\eps=0$ because $\partial_x\partial_y\phi =\partial_y\partial_x\phi$~\cite{justin16}. The leading contributions to $\nabla\cdot \vec J^a$ and $\nabla \cdot \vec J_\eps^a$ will contain products of $\phi$ with either $\delta \beta $ or $\delta \mu$, but these are beyond the accuracy of our linear-response calculation and will henceforth be neglected. On the other hand, as we will show now, $\nabla_j\Pi_{i,j}^a$ will provide a non-trivial contribution to the Euler equation.

For small fields, the drift velocity $\vec u$ is proportional to $\vec E$. When TRS is broken, $\Pi_{i,j}^a$ is zero to first order in $\vec E$ because $\vec \Omega_\vec p$ is even in $\vec p$. The leading term in $\Pi_{i,j}^a$ emerges to first order in $\vec E$ and $\vec u$ and reads,
\bea
\Pi_{i,j}^a &=& \frac{e}{\hbar} \sum_\vec p p_i \epsilon_{jkl} E_k \Omega_l
\lp-\frac{\partial f^0}{\partial \eps}\rp \vec p\cdot \vec u\nn
&=&  u_i \epsilon_{jkl} E_k \mathcal{B}_l,\nonumber \\
 \mathcal{B}_l&=& \mathcal{B}_{i,l}=\frac{e}{\hbar}\sum_\vec p p_i^2 \Omega_l \lp-\frac{\partial f^0}{\partial \eps}\rp.\label{eq:Pia}
\eea
We note that $\mathcal{B}_l$ is related to the Berry curvature at the Fermi surface and is independent of subscript $i$ in the rotationally symmetric 2D system considered in this work $(\mathcal{B}_{x,l}=\mathcal{B}_{y,l})$. Taking the derivative of the anomalous stress tensor, we obtain
\bea
\partial_j\Pi_{i,j}^a &=& \mathcal{B}_l \epsilon_{jkl} \partial_j (u_i E_k) \nn
&=& u_i \lp \nabla\times  \vec E \cdot \boldsymbol{\mathcal{B}}\rp +
\lp \vec E\times \boldsymbol{\mathcal{B}} \cdot \nabla\rp u_i. \label{eq:astress}
\eea
Combining Eqs.~\eqref{eq:enthalpy}, \eqref{eq:np} and \eqref{eq:astress} with Eq.~\eqref{eq:pre-euler}, we obtain the Euler equation for anomalous Hall (AH) materials
\bea
\partial_t (\rho \vec u) +  [\rho (\vec u \cdot \nabla) \vec u+\vec u\nabla\cdot(\rho\vec u)]&&\label{eq:eom}\\
+\vec u \lp \nabla\times  \vec E \cdot \boldsymbol{\mathcal{B}}\rp +
\lp \vec E\times \boldsymbol{\mathcal{B}} \cdot \nabla\rp \vec u &=&  eN \vec E,\nonumber
\eea
where we have introduced the mass density $\rho=W/v^2$. Here, we have assumed that the pressure gradient term $\nabla P$ acts analogously to the electric field and can thus be discarded. In the following, we focus on the steady state, and therefore simplify the Euler equation using $\partial_t \vec u=0$ and $\nabla\times \vec E=0$. Moreover, we assume the electron fluid to be incompressible, which is a good approximation at small $\vec u$. In this case, Eq.~\eqref{eq:cont} leads to $\nabla \cdot \vec u=0$. Finally, we arrive at the steady-state Navier-Stokes equation (NSE)
\be
\rho (\vec u \cdot \nabla)\vec u + (\vec E\times \boldsymbol {\mathcal B}\cdot \nabla) \vec u = eN \vec E +\eta \nabla^2 \vec u. \label{eq:NS}
\ee
where we have added a phenomenological viscous term with a strength $\eta$, arising from dissipative electron-electron interactions in Eq.~\eqref{eq:boltz}~\cite{Briskot2015,Principi2016}.

\section{Poiseuille Flow }
  We apply the NSE of Eq.~\eqref{eq:NS} for the simplest case where electrons are only allowed to move in one direction, i.e., $u_y =0$ and $u_x \equiv u_x(y)$, due to an applied electric field $\vec E =E_x \hat{\vec x}$. We also focus on the case of a 2D AH material where
  $\boldsymbol {\mathcal{B}}=\mathcal {B} \hat{\vec {z}}$. Equation~\eqref{eq:NS} then becomes
  \be
   -E_{x}\mathcal {B}\partial_y  u_x = e N   E_x  + \eta \partial_y^2  u_x.\label{eq:poiseuille}
   \ee
   The convective term $(\vec u \cdot \nabla)\vec u$ disappears because $\partial_xu_x=0$. We make Eq.~\eqref{eq:poiseuille} dimensionless by defining  $\tilde y = y/w$, where $w$ is the width of the 1D channel. Thus we obtain
  \be
  -b \partial_{\tilde y} \tilde u_x = 1 +  \partial_{\tilde y}^2  \tilde u_x,
  \ee
  where
  \be
  b= \frac{w \mathcal {B} E_x}{\eta},\quad \tilde u_x=u_x/u_0, \quad u_0 = \frac{w^2 eN E_x}{\eta }
  \ee
  Applying no-slip boundary conditions $\tilde u_x (0) = \tilde u_x (1)=0$, the solution of this ordinary differential equation becomes
  \be
  \tilde u _x (\tilde y)= \frac{e^{b (1-\tilde y)}-e^{b}(1-\tilde y)-\tilde y}{b (1-e^{b})}\label{eq:sol1}.
  \ee
   In the limit $b\to 0$, we recover from Eq.~\eqref{eq:poiseuille} the familiar Poiseuille flow profile, where $\tilde u=\tilde y(1-\tilde y)/2$, i.e., a parabolic profile with the highest velocity at the center at $y=w/2$. However, when the Berry curvature is nonzero, the velocity profile deviates from the ordinary Poiseuille result and the location of the velocity maximum is controlled by the size and sign of $\mathcal {B}$. We will see below how Berry curvature modifies the Poiseuille flow in a concrete model.

In order to observe the Poiseuille flow in AH materials, the magnitude of the
viscous longitudinal current $J_x=eNu_x$ must exceed the Hall
current $J_y=(e^2/h) \mathcal{C} E_x$, where
$\mathcal{C}=(2\pi)^{-1}\sum_{\pm}\int d^2\vec p\ \Omega_\vec
p^{\pm}f^0$ is the Berry flux. We use a gapped Dirac model with
$\Delta =100$ meV. The Fermi energy $\mu=\Delta$ is set to the bottom
of the conduction band at $100 \text{ K}$ to suppress the contribution of
possible chiral edge states in the gap. Using these parameters,
we obtain $N = 2\times 10^{10}\ \rm cm^{-2}$,
$\mathcal{B}/e=0.005\ \rm fs/nm^2$, $\mathcal{C}=0.47$ (note that gapped Dirac bands have Chern numbers $\pm 1/2$ for the valence and conduction bands, respectively). In Fig.~\ref{Fig1}(a), the Hall
current $J_y$ is linearly proportional to the electric field while the
longitudinal current $J_x$ at the center $y=w/2$ is initially
proportional to $E_x$ and then saturates for large $E_x$.  Importantly, at
very small electric fields, the viscous longitudinal current $J_x$
dominates over the Hall current $J_y$. The saturation of $J_x$ can be
seen from Eq.~\eqref{eq:poiseuille}: when the $E_x$ term becomes
much larger than the viscous term, $u_x$ becomes independent of $E_x$.

In the regime where $J_x$ exceeds $J_y$, there exists a window
of size $\propto E_x$ where the Berry curvature dramatically modifies the
Poiseuille profile as shown in Fig.~\ref{Fig1}(b). We vary $E_x$ in
Eq.~\eqref{eq:sol1} and show that at intermediate $E_x$, the maximum
velocity departs from the center to a position controllable by $E_x$
[see Fig.~\ref{Fig1}(c)]. This asymmetric Poiseuille flow can be observed with the state-of-the-art methods such as scanning-probe microscopy based on nitrogen-vacancy centers~\cite{yacoby2020}.

  \begin{figure}
\centering
\includegraphics{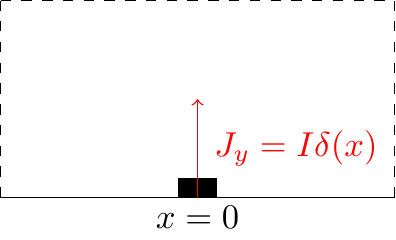}
\caption{The half-plane geometry with a single contact.\label{Fig0}}
\end{figure}
\begin{figure*}
  \centering
\includegraphics[height=6cm]{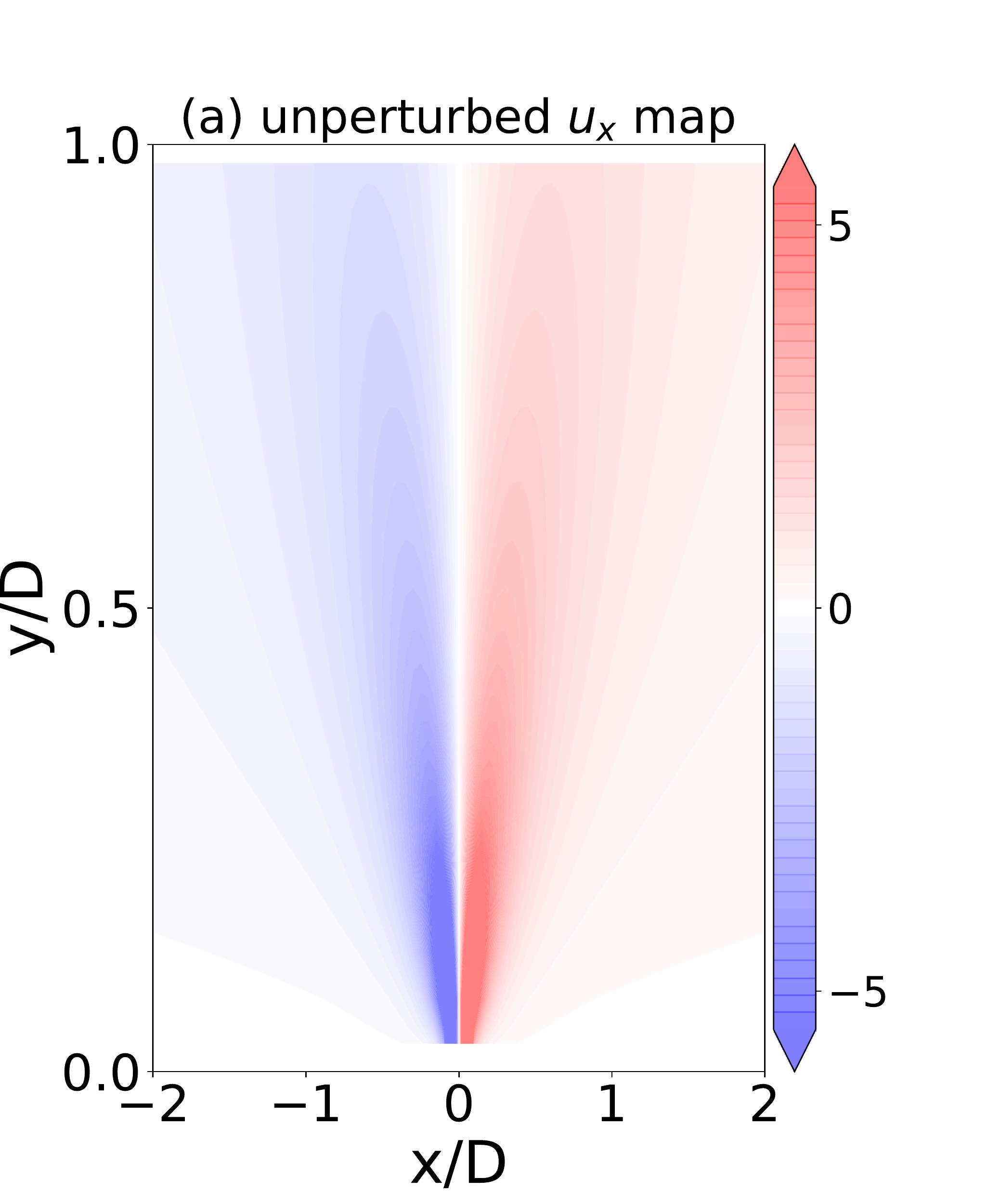}
\includegraphics[height=6cm]{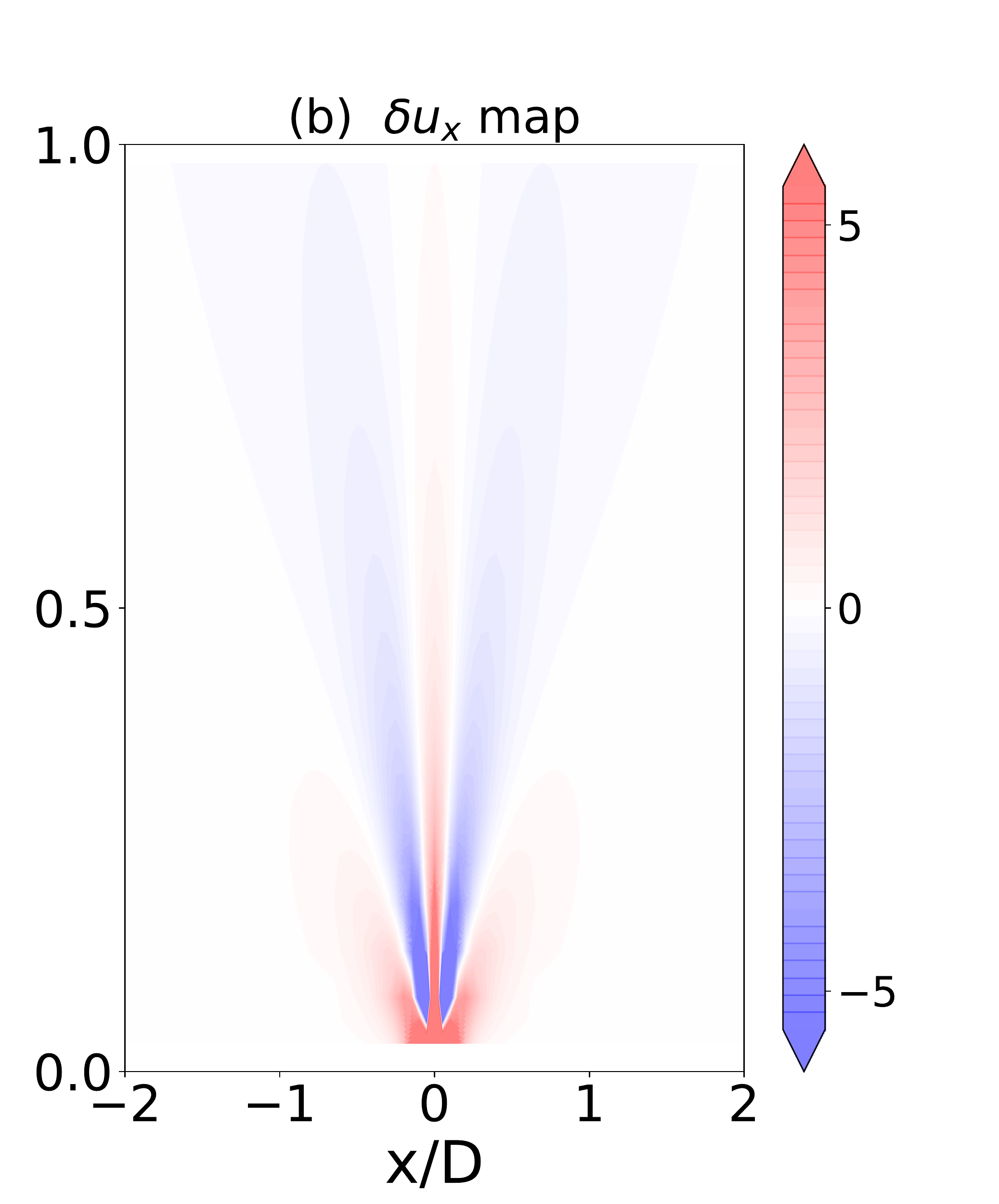}
\includegraphics[height=6cm]{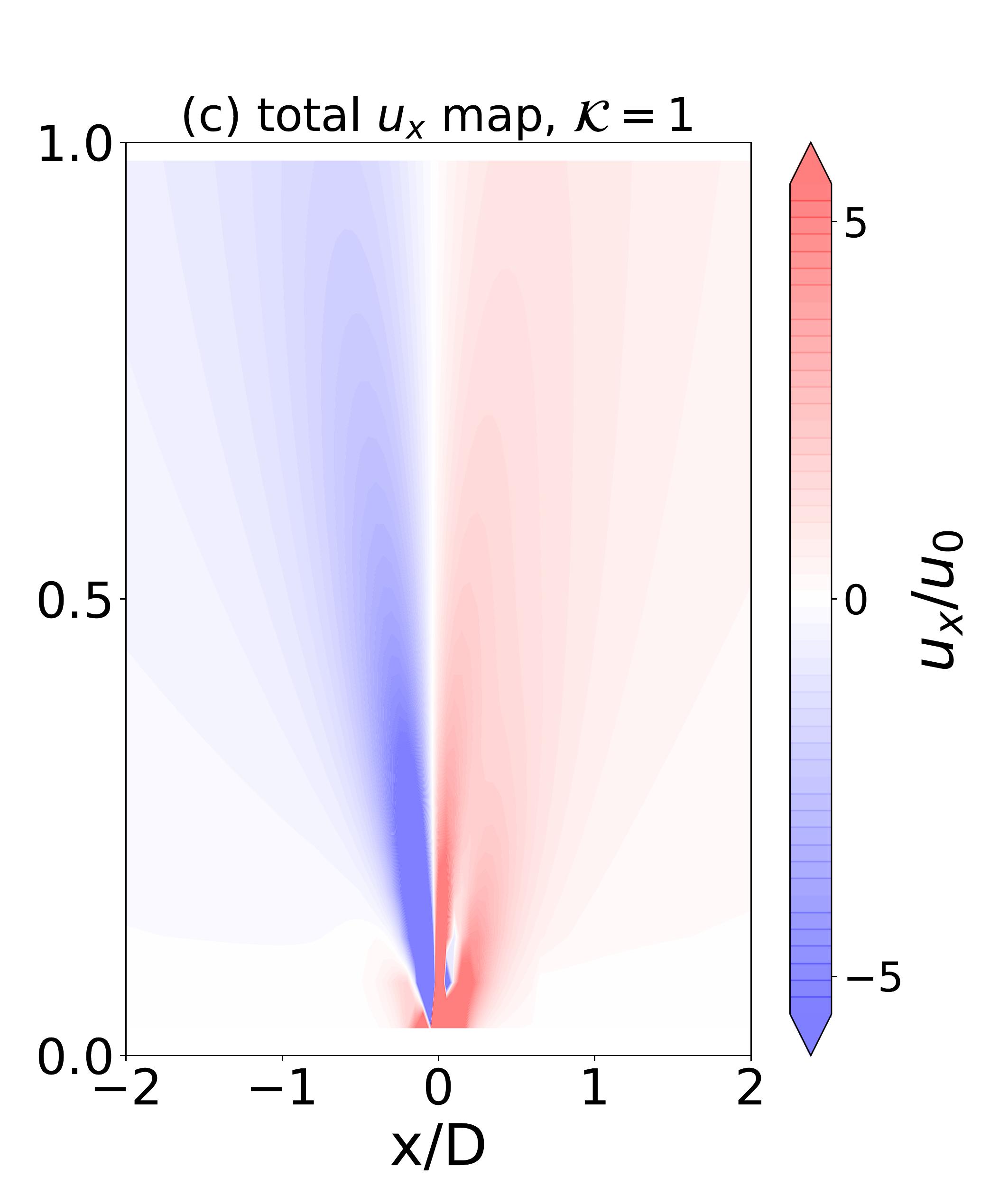}
\caption{(a) Unperturbed solution $u_x^\zero$, (b) correction $\delta u_x$ and (c) total velocity $u_x$ after Fourier transformation. We set $\mathcal{K}=1$.\label{Figux}}
\end{figure*}

   \begin{figure*}
     \centering
\includegraphics[height=6cm]{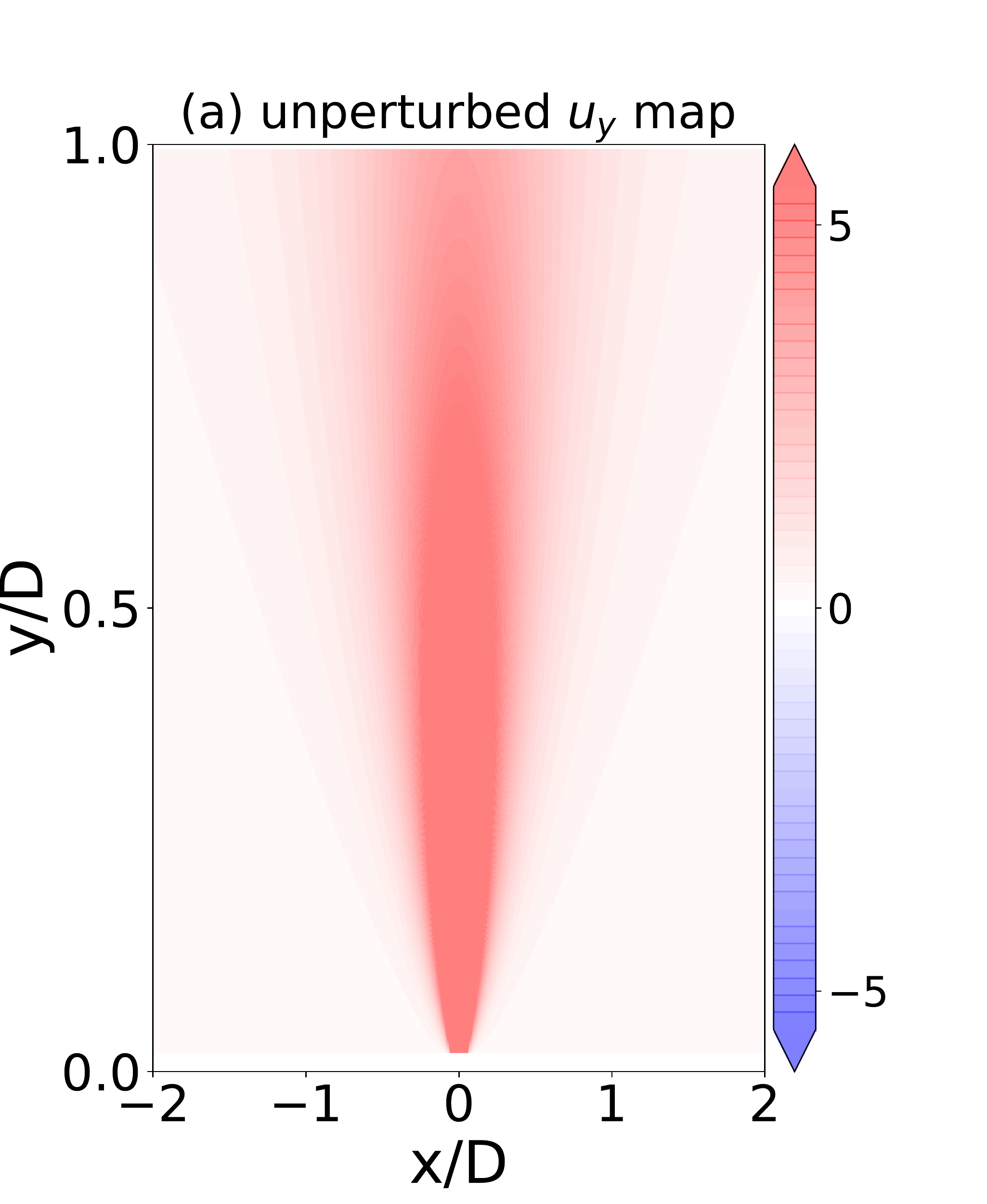}
\includegraphics[height=6cm]{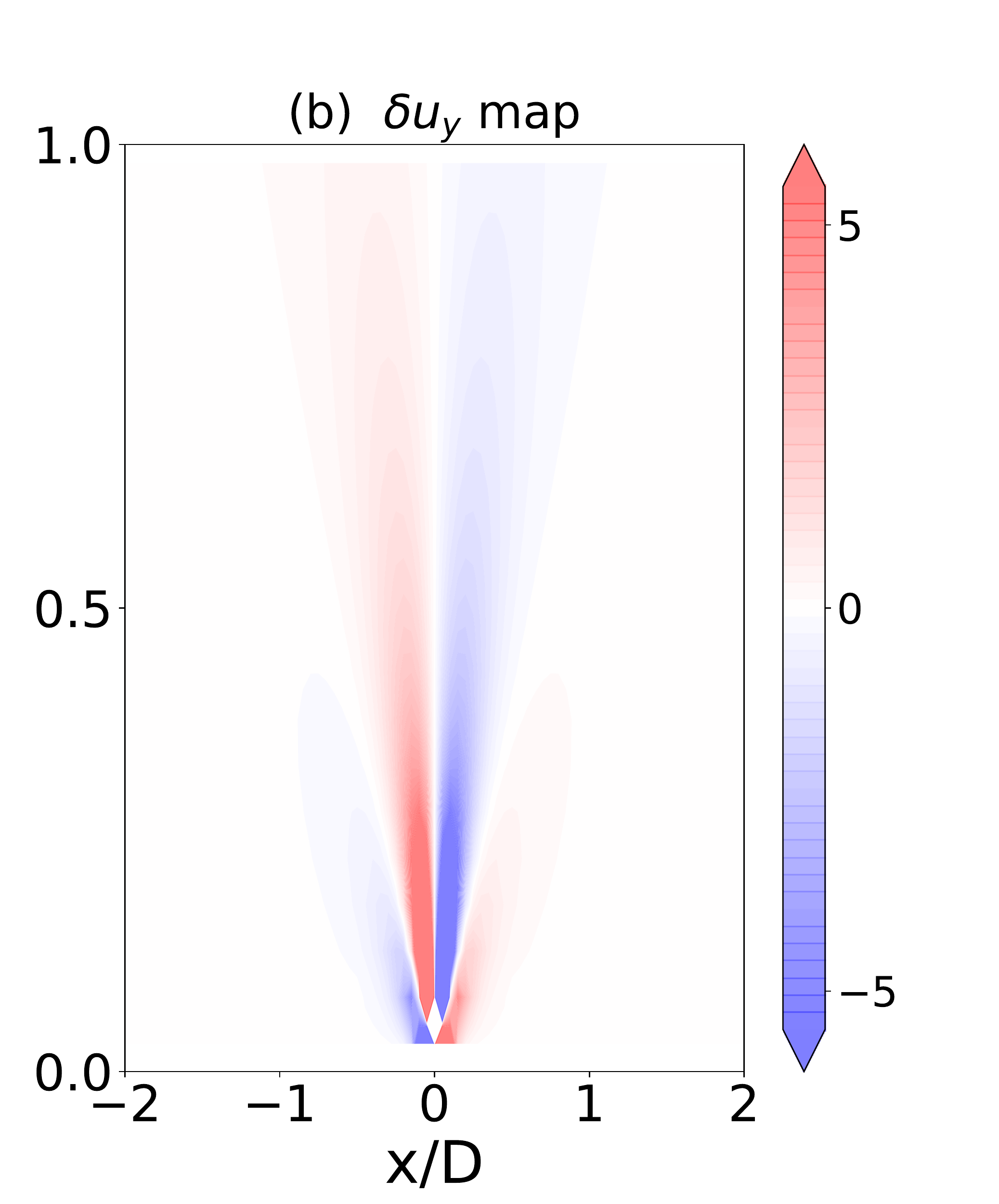}
\includegraphics[height=6cm]{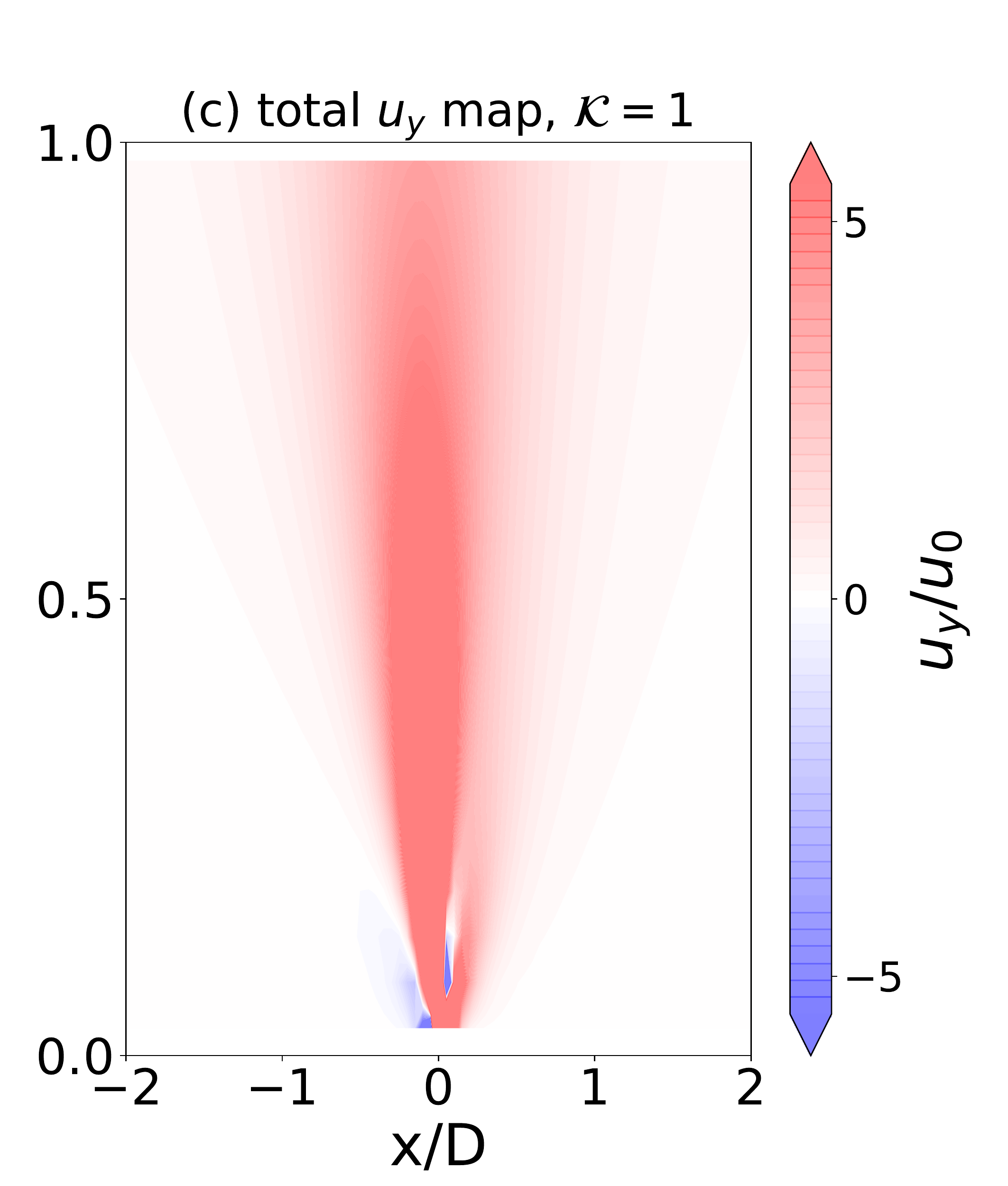}
\caption{(a) Unperturbed solution $u_y^\zero$, (b) correction $\delta u_y$ and (c) total velocity $u_y$ after Fourier transformation. We set $\mathcal{K}=1$.\label{Figuy}}
 \end{figure*}

\section{Berry curvature and vorticity} We shift our attention
   now to 2D geometries. We can obtain the vorticity field $\boldsymbol{
     \omega}=\nabla\times \vec u$ from the NSE by taking the curl on
   both sides of the Eq.~\eqref{eq:NS}. The result is,
\be
   (\vec E \times \boldsymbol {\mathcal B}\cdot \nabla)  \omega = \eta \nabla^2  \omega \label{eq:vorticity},
  \ee
  where $\boldsymbol{\omega}=\omega \hat{\vec z}$ because the system
  is confined to two dimensions. This equation implies that the Berry
  curvature contributes in a non-trivial way to the vorticity. However, we should point out that
  a nonzero vorticity $\boldsymbol{\omega}$ is not equivalent to the existence of current whirlpools. For
  instance, in previous works it was shown that, despite $\boldsymbol{\omega}$ being nonzero for viscous electrons, no current whirlpool emerge in a
  semi-infinite 2D geometry~\cite{Pellegrino2016, Falkovich2017}. In
  contrast, it was found that the whirlpools solely arise from current
  backflow in a finite geometry. However, as we will show below, the
  nontrivial vorticity in Eq.~\eqref{eq:vorticity} will give rise to
  current whirlpools even in the absence of boundaries.

  Turning to the analysis of the 2D flow, we apply Eq.~\eqref{eq:NS} in a half-plane geometry with an injected current at the origin ($x=y=0$) as shown in Fig.~\ref{Fig0}. Here we assume a flow at low Reynolds number or small $\vec u$ such that we can neglect the convective term $(\vec u\cdot\nabla)\vec u$ as well as the Hall current.
   We express Eq.~\eqref{eq:NS} in units of acceleration by dividing both sides by the mass density $mN$,
   \be
   \frac{1}{mN} (-\nabla \phi\times \boldsymbol {\mathcal B}\cdot \nabla) \vec u = -\frac{e}{m}\nabla \phi +\nu   \nabla^2 \vec u-\frac{\vec u}{\tau} , \label{eq:NS-3}
   \ee
   where we have introduced the kinematic viscosity $\nu = \eta/\rho=\eta/(m N)$. We have also added a relaxation time $\tau$ due to momentum-relaxing scattering processes (electron-phonon or electron-impurity) that allows us to define a length scale $D =\sqrt{\nu \tau}$ known as the diffusion length. Suppose we apply a current density $J_y=I \delta(\vec r)$ at the origin. We can scale the quantities to get a dimensionless equation
   \be
   \mathcal{K} (-\tilde\nabla\tilde\phi\times\hat{\vec z}\cdot\tilde\nabla)\tilde {\vec u} =-\tilde \nabla \tilde \phi + \tilde\nabla^2\tilde{\vec u}-\tilde{\vec u},\label{eq:NSnodim}
   \ee
   where  $\tilde \phi = \phi/\phi_0$, $\phi_0= Im/Ne^2\tau$, $\tilde u=u/u_0$, $u_0=I/NeD$, $(\tilde x,\tilde y)=(x/D,y/D)$ and $\mathcal{K}=I\Delta^2b(\mu,T)/(eD^2N^2hv^2k_BT)$. To get $\mathcal{K}$, we have used $\boldsymbol{\mathcal{B}}= \mathcal{B}_z \hat{\vec z}$ where $\mathcal{B}_z=e\Delta^2 b(\mu,T)/hv^2k_BT$ and $b(\mu,T)$ is a dimensionless integral of Eq.~\eqref{eq:Pia} [see Eq.~\eqref{eq:bz} in Appendix]. Hereafter, we drop all the ``tilde'' accents for better readability while keeping dimensionless quantities.

   Equation~\eqref{eq:NSnodim} is nonlinear because of the coupling between $\phi$ and $\vec u$ on the left-hand side (LHS). Using perturbation theory in this coupling, we can linearize this equation by introducing
     \be
     \vec u = \vec u^{(0)} +\mathcal{K} \delta\vec u,\quad\quad
     \phi= \phi^{(0)} + \mathcal{K} \delta\phi.\label{eq:total}
     \ee
     Comparing terms in Eq.~\eqref{eq:NSnodim}, we obtain the following equations for the unperturbed result and the perturbation,
     \bea
     &&-\nabla\phi^\zero+ \nabla^2\vec u^\zero-\vec u^\zero= 0\label{eq:zero},\\
     &&-\nabla\delta\phi+ \nabla^2\delta \vec u-\delta\vec u=\lp-\nabla\phi^\zero\times\hat{\vec z}\cdot\nabla \rp \vec u^\zero. \label{eq:delta}
     \eea
 The incompressibility condition $\nabla\cdot \vec u=0$ leads to the Laplace equation $\nabla^2\phi^\zero=0$ for the unperturbed electric potential. In contrast, the perturbation satisfies a Poisson equation with $\nabla^2\delta\phi\neq 0$ due to the right-hand side of \eqref{eq:delta}, implying the presence of an induced charge by the Berry curvature. As we show later, for the half-plane geometry, $\nabla^2\delta\phi$ from Eq.~\eqref{eq:delta} displays an electric dipole originating from the coupling of electric field and drift velocity induced by the Berry curvature [see inset of Fig.~\ref{Figphi}(b)].

     This half-plane geometry preserves translational invariance in the $x$ direction, so we seek solutions in the form of Fourier transforms,
   \be
   \vec u (x,y)=\int dk \vec u_k(y) e^{ikx},\qquad
   \phi (x,y) = \int dk \phi_k(y) e^{ikx}. \label{eq:FT}
   \ee
   The solutions of Eq.~\eqref{eq:zero} follow from Ref.~\onlinecite{Pellegrino2016}. To be specific, we use no-slip boundary conditions where $u_x^\zero$ vanishes for $y=0$ and $u_y^\zero(y=0)=\delta(x)$ arising from the injected current at the origin. We obtain (see Appendix):
     \bea
     u_{k,x}^{(0)}(y)&=& \frac{i kq}{|k|(|k|-q)}\lp  e^{-|k|y}- e^{-qy}\rp,\nn
     u_{k,y}^{(0)}(y)&=&-\frac{1}{(|k|-q)} \lp q e^{-|k|y}- |k| e^{-qy} \rp,\label{eq:zerosol}\\
     \phi_k^{(0)}(y) &=& -\frac{1}{|k|} \frac{q}{|k|-q} e^{-|k|y},\nonumber
     \eea
     where $q=\sqrt{k^2+1}$.

    The solutions of the first-order correction, $\delta\vec u$ and $\delta \phi$ in Eq.~\eqref{eq:delta}, consist of homogeneous and inhomogeneous solutions. The homogeneous solutions obey the same equation as the unperturbed solution, but with different boundary conditions. The complete first-order corrections have the form,
     \begin{align}
   \delta u_{k,x}(y) &=
      \frac{-a_1i}{k} e^{-|k| y} - \frac{a_3}{q}e^{-qy}+\delta u_{x,1}e^{-(q+|k|)y},\nn
   \delta u_{k,y}(y) &=
     \frac{a_1 }{|k|} e^{-|k| y} - \frac{a_3 ik}{q^2} e^{-qy}+ \delta u_{y,1}e^{-(q+|k|)y},\nn
   \delta \phi_k(y) &=
     \frac{a_1}{k^2}  e^{-|k| y} + \delta \phi_{1}e^{-(q+|k|)y},\label{eq:deltasol}
     \end{align}
where the inhomogeneous solutions are,
\bea
   \delta u_{x,1}(y)& =& \frac{q^3+2q^2|k|-|k|^3}{(q-|k|)(5k^2q+2(k^2+q^2)|k|},\label{eq:soldelta}\\
    \delta \phi_1(y) &=& \frac{i k|k|}{q^2+q|k|-2k^2},\nonumber
   \eea
   as well as $\delta u_{y,1}=(ik/(|k|+q))\delta u_{x,1}$ which follows from $\nabla\cdot\delta \vec u=0$. Imposing boundary conditions where $\delta \vec u$ vanishes at the boundary $\delta u_{k,x}(y=0)=\delta u_{k,y}(y=0)=0$, we obtain the coefficients
 \bea
 a_{1} &=& \frac{i k^3 (-k^2+q^2+q|k|)}{(q-|k|)^2[5k^2q+2|k|(k^2+q^2)]},\nn
  a_{3} &=& \frac{q^3 (-k^2+q^2+q|k|)}{(q-|k|)^2[5k^2q+2|k|(k^2+q^2)]}.\label{eq:coef}
  \eea
  We note that $a_1$ is anti-symmetric as a function of $\vec k$ while $a_3$ is symmetric. As a result, $\delta u_{k,x}$ is symmetric while $\delta u_{k,y}$ and $\delta \phi_k$ are anti-symmetric as functions of $\vec k$. Importantly, the unperturbed counterparts of these functions have the opposite symmetry: $u_{k,x}^\zero$ is anti-symmetric while $u_{k,y}^\zero$ and $\phi_k^{\zero}$ are symmetric.

Combining the unperturbed solution in Eq.~\eqref{eq:zerosol} with the first-order correction in Eq.~\eqref{eq:deltasol}, we can obtain the velocity and potential profile in real space via a numerical Fourier transformation using Eq.~\eqref{eq:FT}.
\begin{figure*}
  \centering
  \includegraphics[height=6cm]{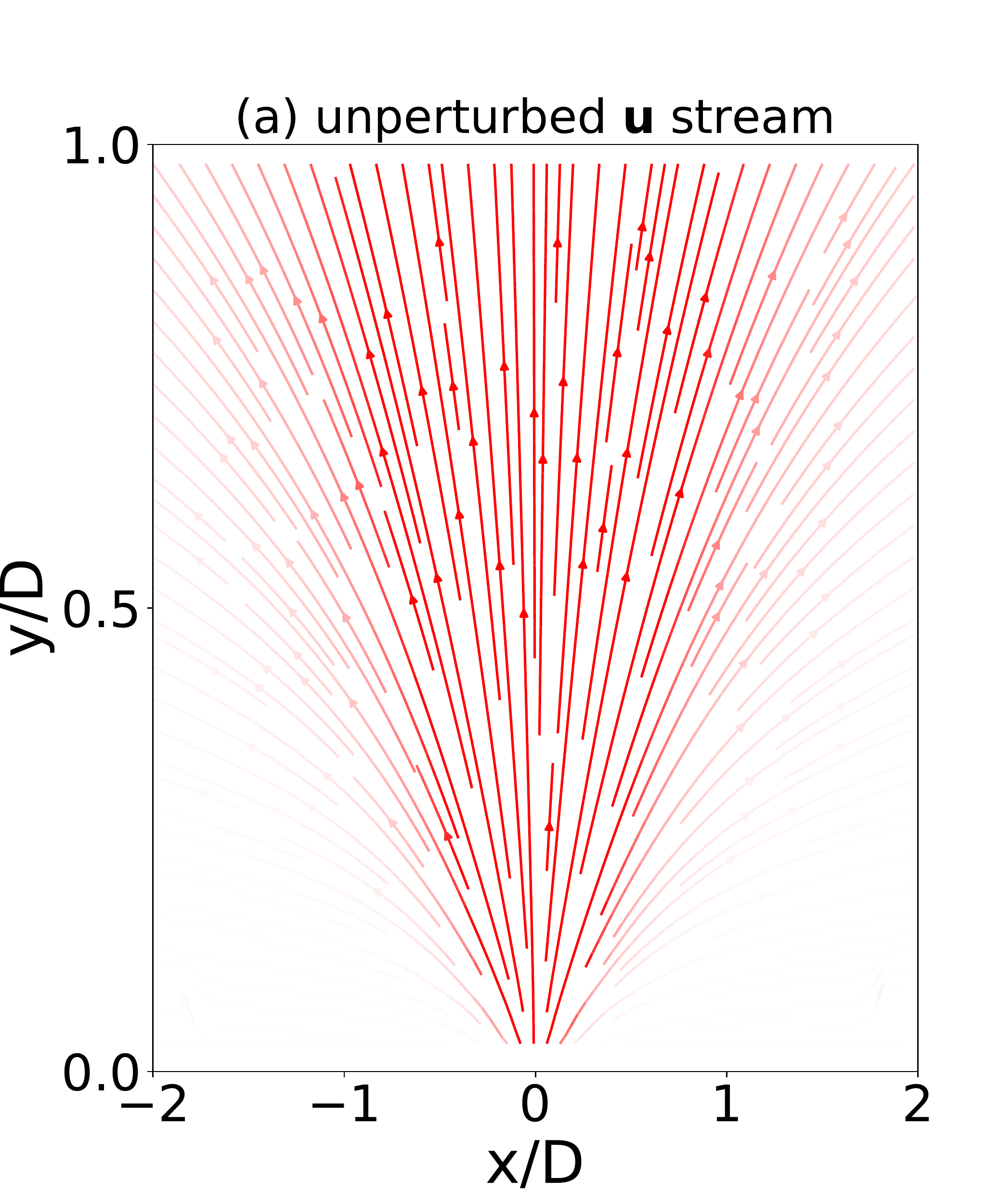}
    \includegraphics[height=6cm]{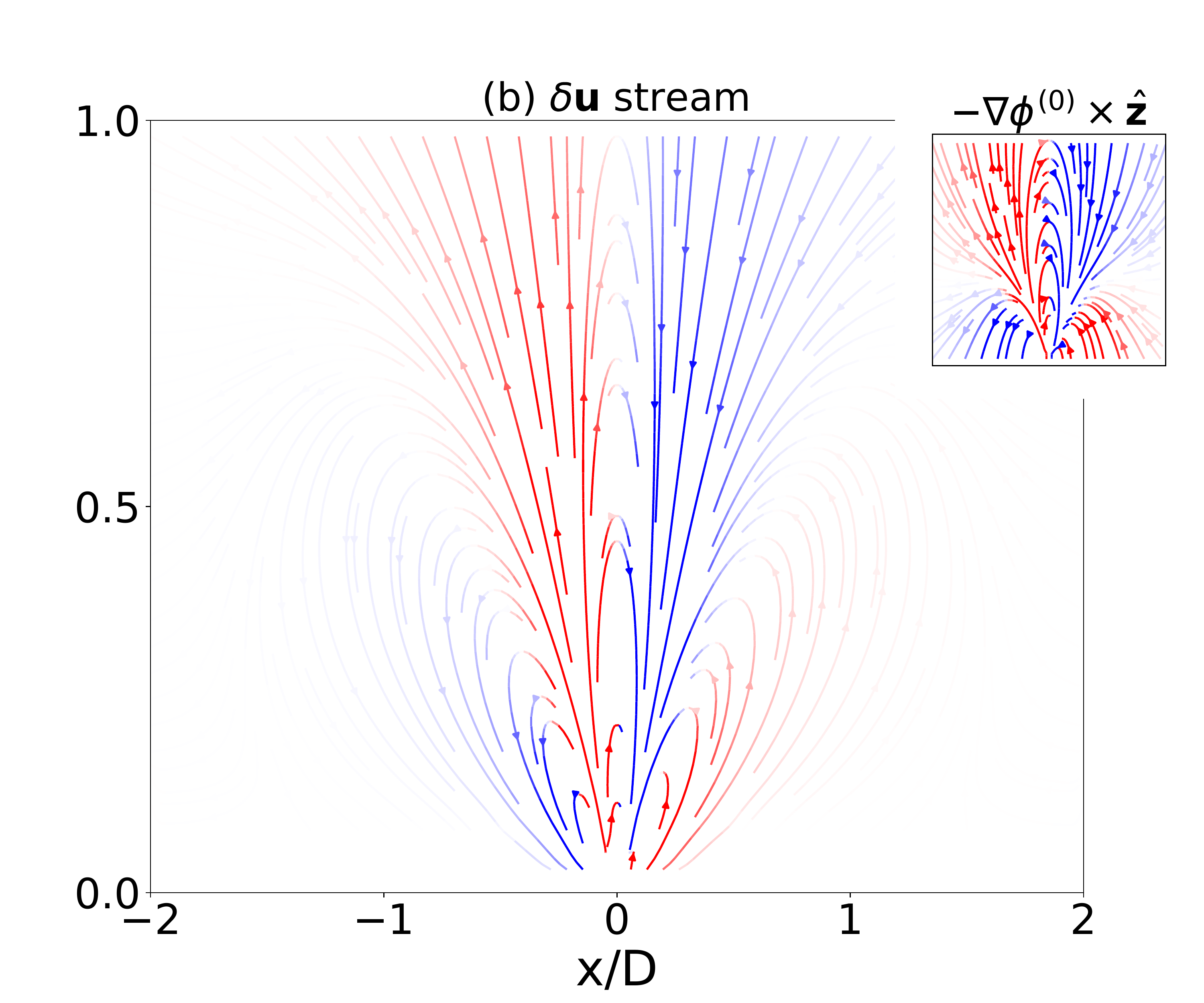}
\includegraphics[height=6cm]{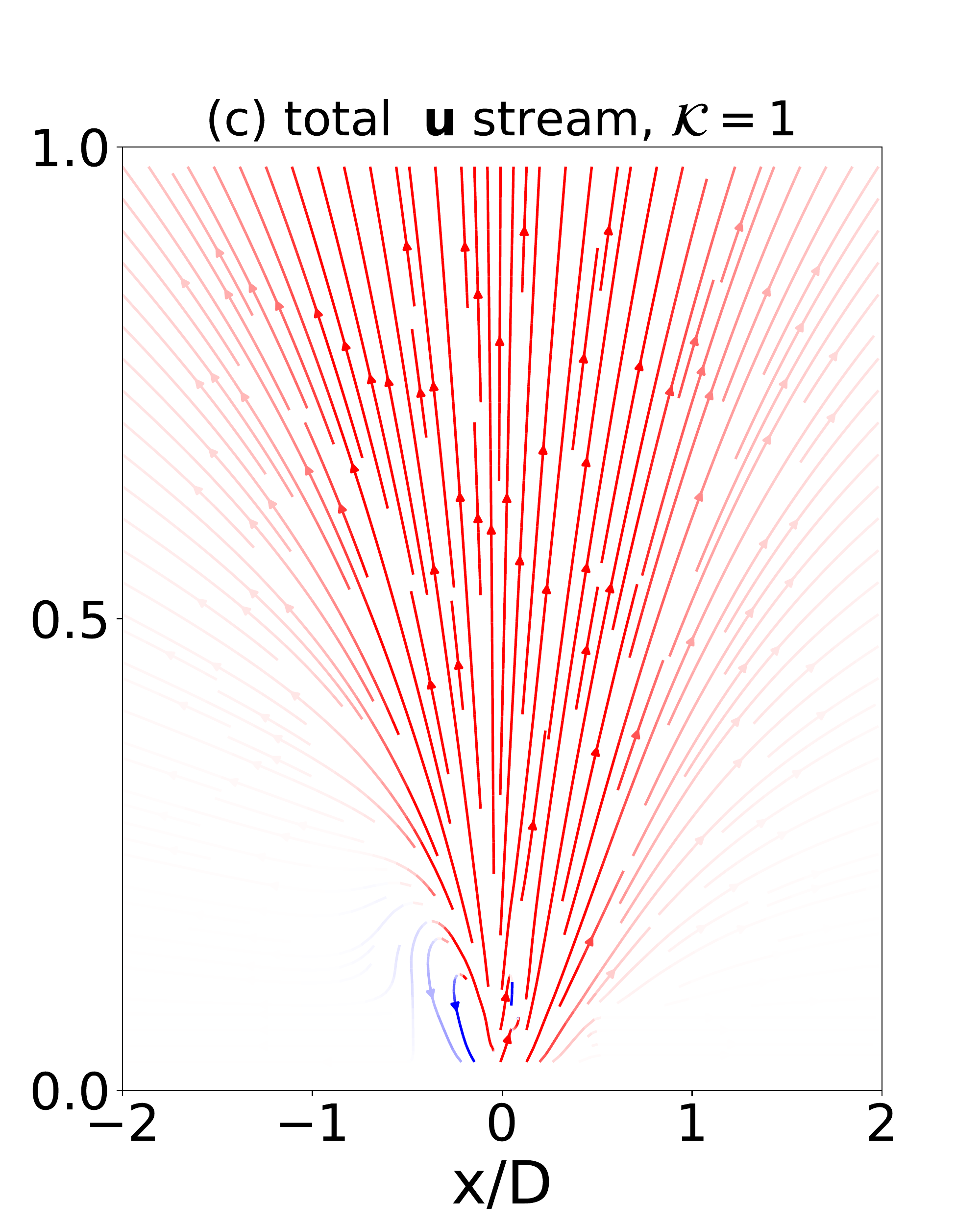}
\caption{ Stream plots of (a) the unperturbed flow profile $\vec u^\zero$, (b) the perturbation $\delta \vec u$ and (c) the total solution $\vec u$. The inset shows the stream of $-\nabla\phi^\zero\times\hat{\vec z}$ with the same $x$ and $y$ scales as the main figure. We set $\mathcal {K}=1$.\label{Figstream} }
\end{figure*}
\begin{figure*}
 \centering
\includegraphics[height=6cm]{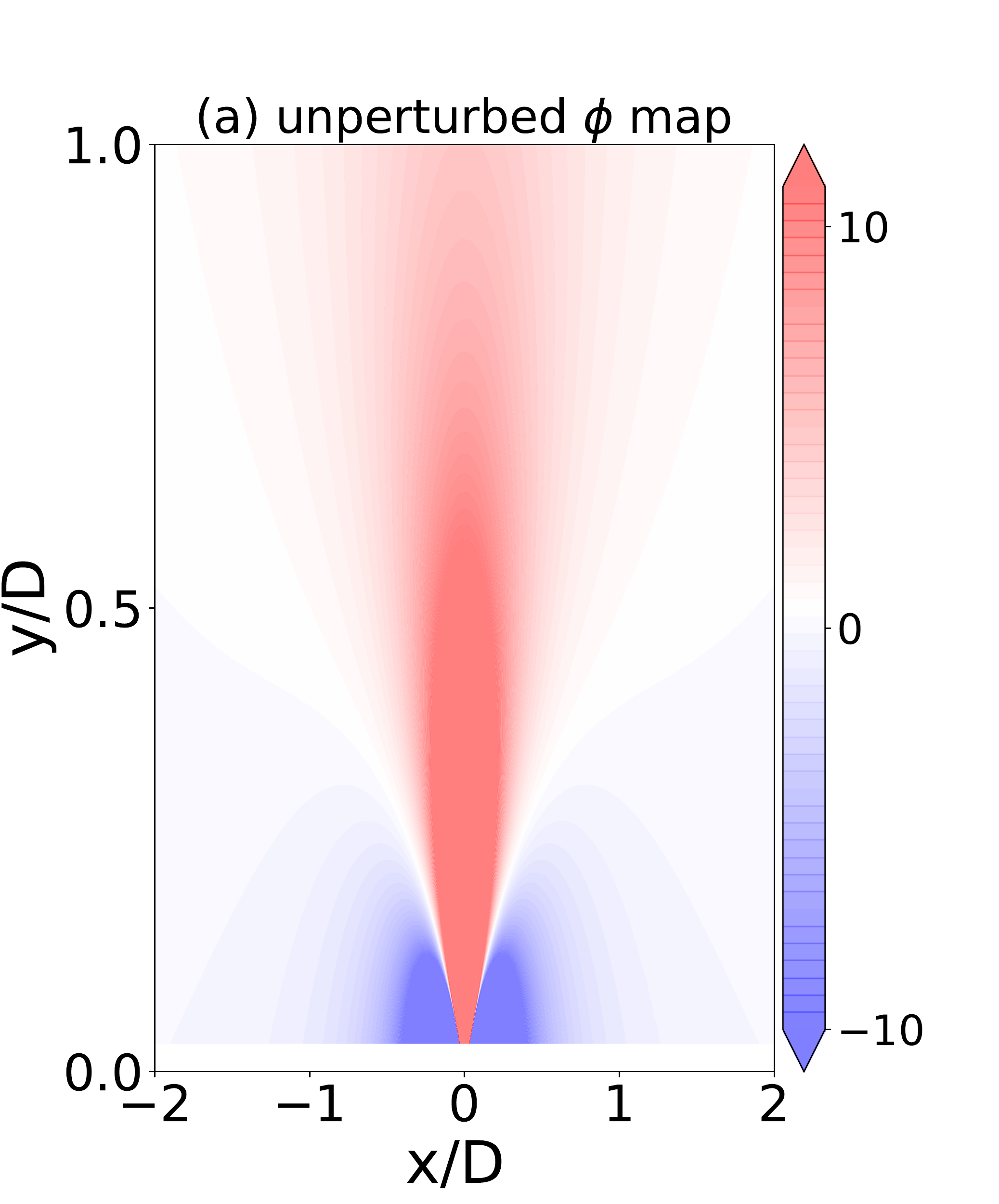}
\includegraphics[height=6cm]{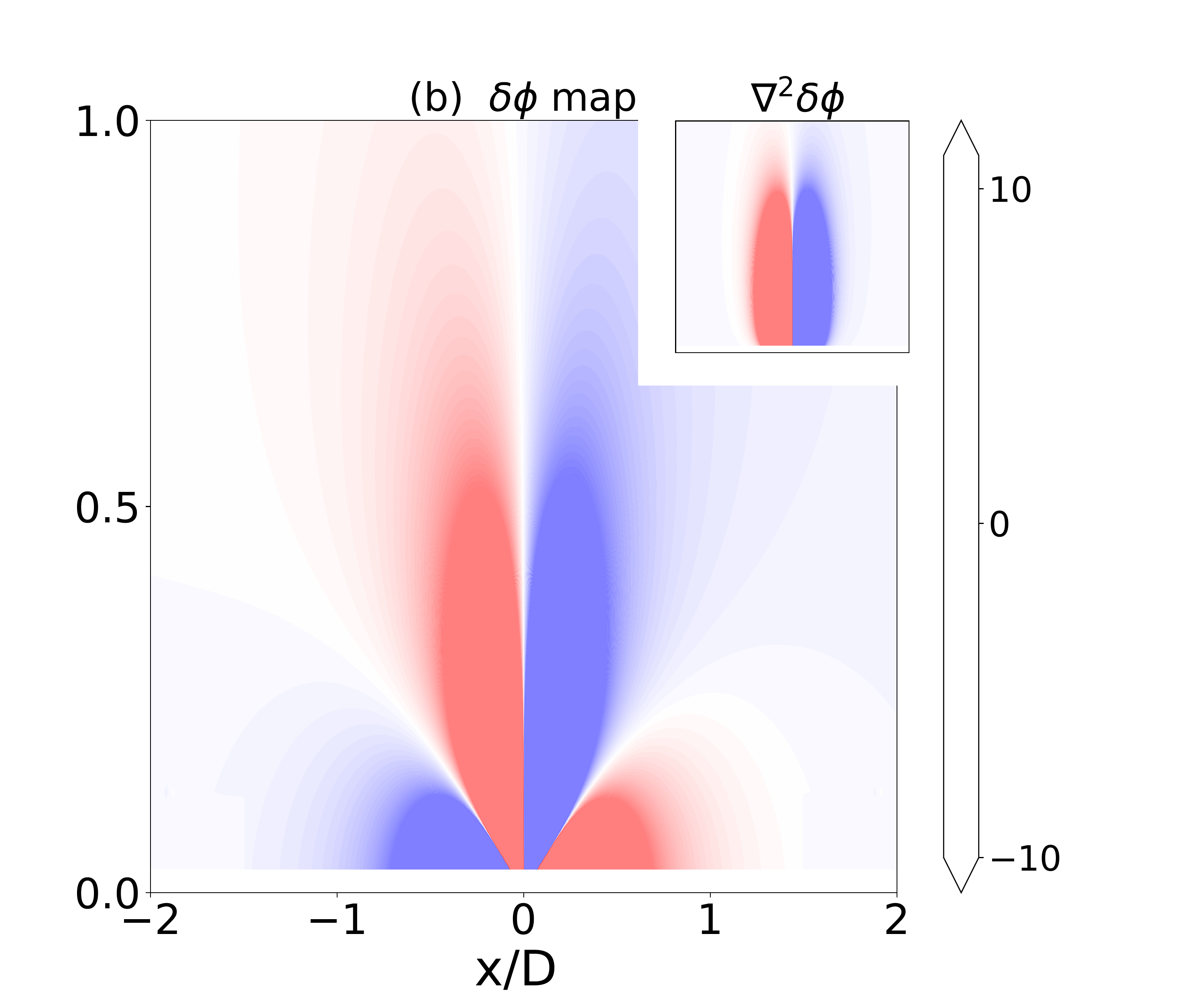}
\includegraphics[height=6cm]{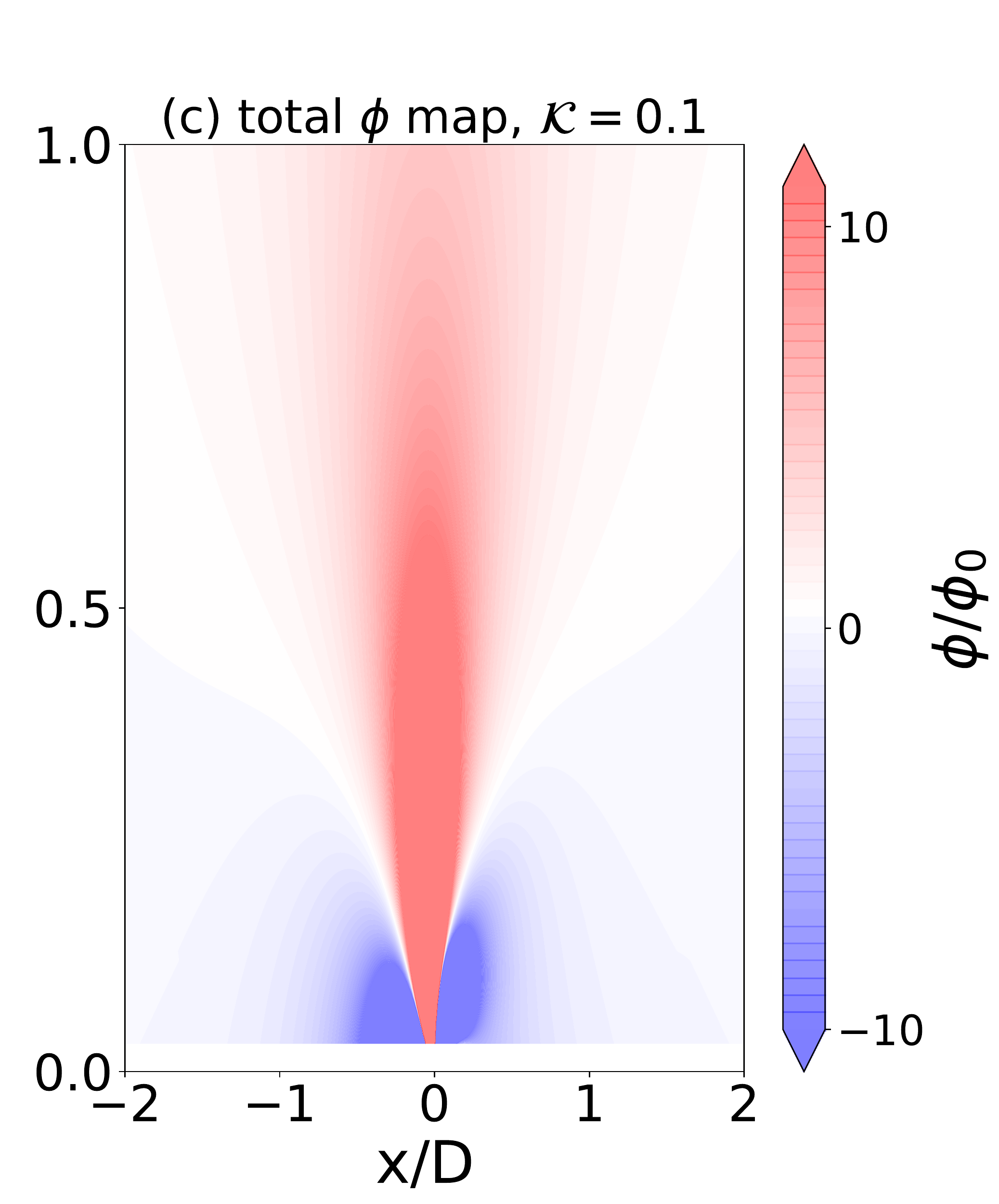}
\caption{(a) Unperturbed solution $\phi^{(0)}$, (b) the correction $\delta \phi$, and (c) the total potential $\phi$ after Fourier transformation. The inset shows $\nabla^2\delta \phi$ with the same $x$ and $y$ scales as the main figure. We set $\mathcal{K}=0.1$.\label{Figphi} }
\end{figure*}
    \begin{figure}
  \centering
  \includegraphics[height=4cm]{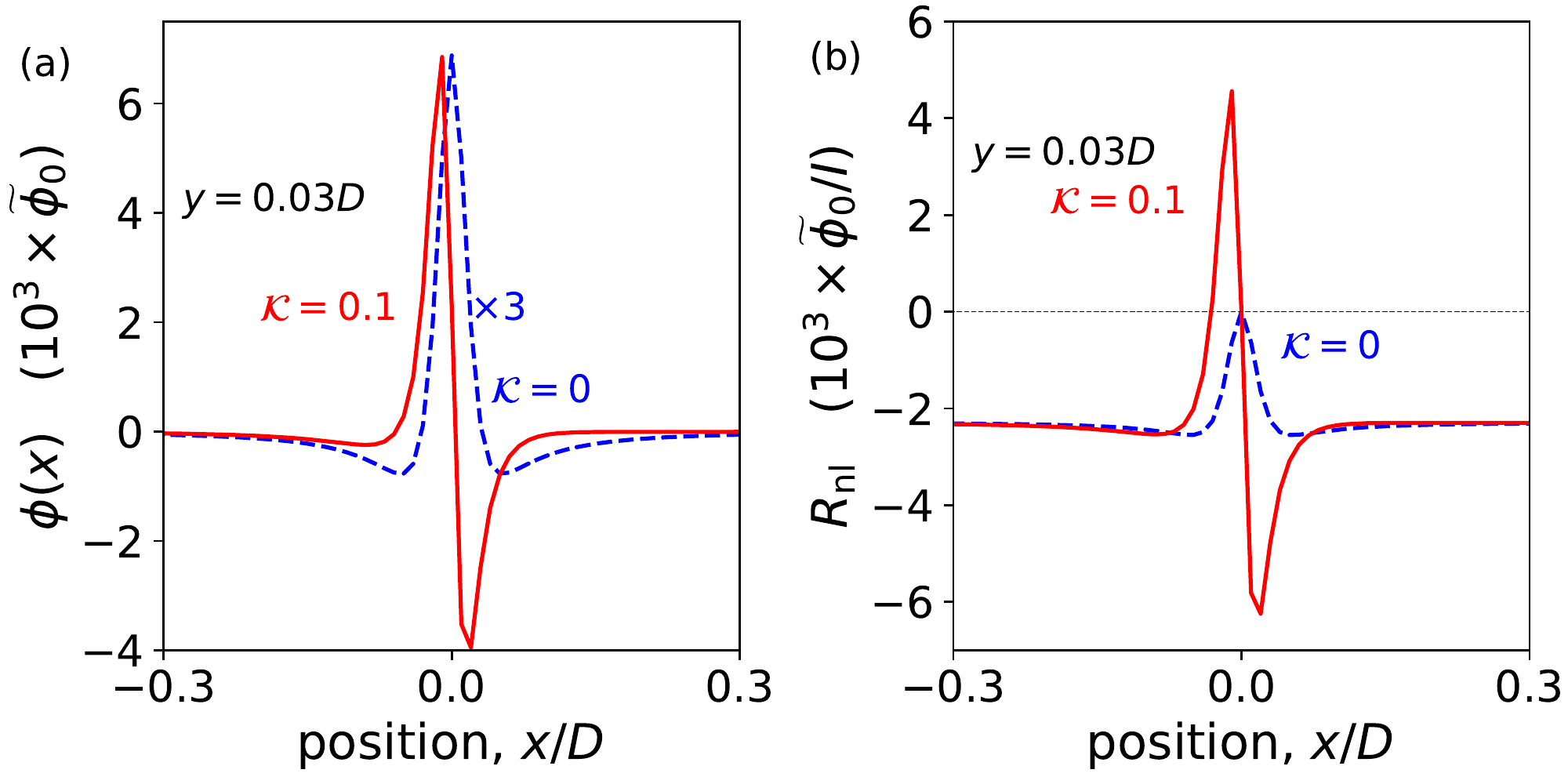}
\caption{ (a) The electric potential and (b) the non-local resistance profile without Berry curvature (blue) and with Berry curvature (red), $\mathcal{K} =0.1$.\label{Fignonlocal} }
    \end{figure}

In Figs.~\ref{Figux} and \ref{Figuy}, we display the components of $\vec
u^\zero$, $\delta \vec u$ and of the full solution $\vec u$. Inheriting the symmetry from its Fourier components, $u^\zero_x$
($u^\zero_y$) is anti-symmetric (symmetric) in $x$, whereas
$\delta u_x$ ($\delta u_y$) is symmetric (anti-symmetric). The
opposite symmetries of the unperturbed solution and the perturbation produce asymmetric flow structures in
the full solution $\vec u$ due to constructive and destructive superpositions.
Far from the inlet, $\vec u$ is dominated by $\vec u^\zero$, while near the inlet, $\vec u$ is strongly modified by the perturbation $\delta\vec u$. Importantly, the values of $u_y$ are not homogeneously positive but can also become negative near the inlet. This indicates non-trivial backflow due to the vorticity generated by the Berry curvature~\eqref{eq:vorticity}.

A stream plot of $\vec u^\zero$ is shown in Fig.~\ref{Figstream}(a). The
color follows the value of $u_y$: red (blue) indicates a positive
(negative) value. The light red stream lines are an order of
magnitude weaker than the main stream lines. The unperturbed flow $\vec u^\zero$ spreads out
from the inlet as expected~\cite{Falkovich2017}. Although no current whirlpool is present, the vorticity $\boldsymbol \omega=\nabla\times \vec u^\zero$ is non-zero as indicated by the curved stream lines away from $x=0$. The unperturbed vorticity $\boldsymbol \omega$ changes sign at $x=0$. Interestingly, the $\delta \vec u$ stream shows an intricate flow pattern where the sign of the vorticity changes twice along azimuthal direction (up-down-up), see Fig.~\ref{Figstream}(b). This complex structure is related to the coupling of the Berry curvature $\boldsymbol {\mathcal{B}}\propto \hat{\vec z}$ to the electric field $-\nabla \phi^{(0)}$ as shown in the inset. The $-\nabla\phi^\zero\times \hat{\vec z}$ profile is rather insensitive to the choice of boundary conditions, and the vortices in $\delta \vec u$ will persist even if we impose no-stress
boundary conditions. In Fig.~\ref{Figstream}(c), we show the resulting $\vec u$ stream for $\mathcal {K} = 1$. Away from the inlet, $\vec u$ approaches $\vec u^\zero$. In the vicinity of the inlet, the flow is highly asymmetric and exhibits whirlpools as a remnant of the contribution from $\delta \vec u$.

If the applied current and the Berry curvature $\mathcal{B}_z$ in experimental setups are small, they might result only in small values of $\mathcal{K}$. Therefore, even if whirlpools may exist, observing them might be challenging. Fortunately, the profile of the electric potential is far more sensitive to the presence of Berry curvature and one can therefore utilize it as an experimental tool to study Berry curvature effects in electron hydrodynamics.
\begin{figure}
\centering
\includegraphics[trim=0cm 0cm 0cm 0cm, clip,width=0.23\textwidth]{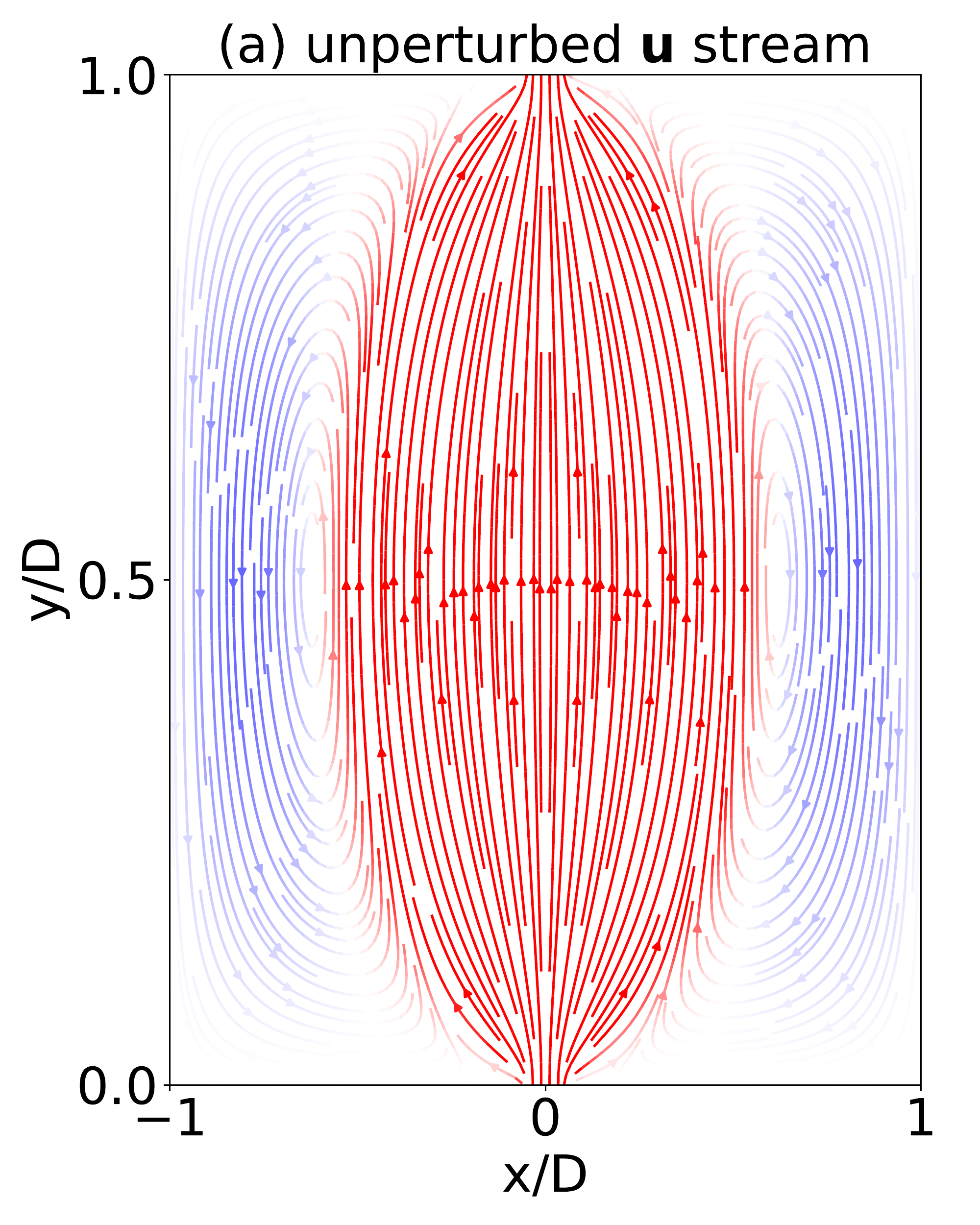}
\includegraphics[trim=0cm 0cm 0cm 0cm, clip,width=0.23\textwidth]{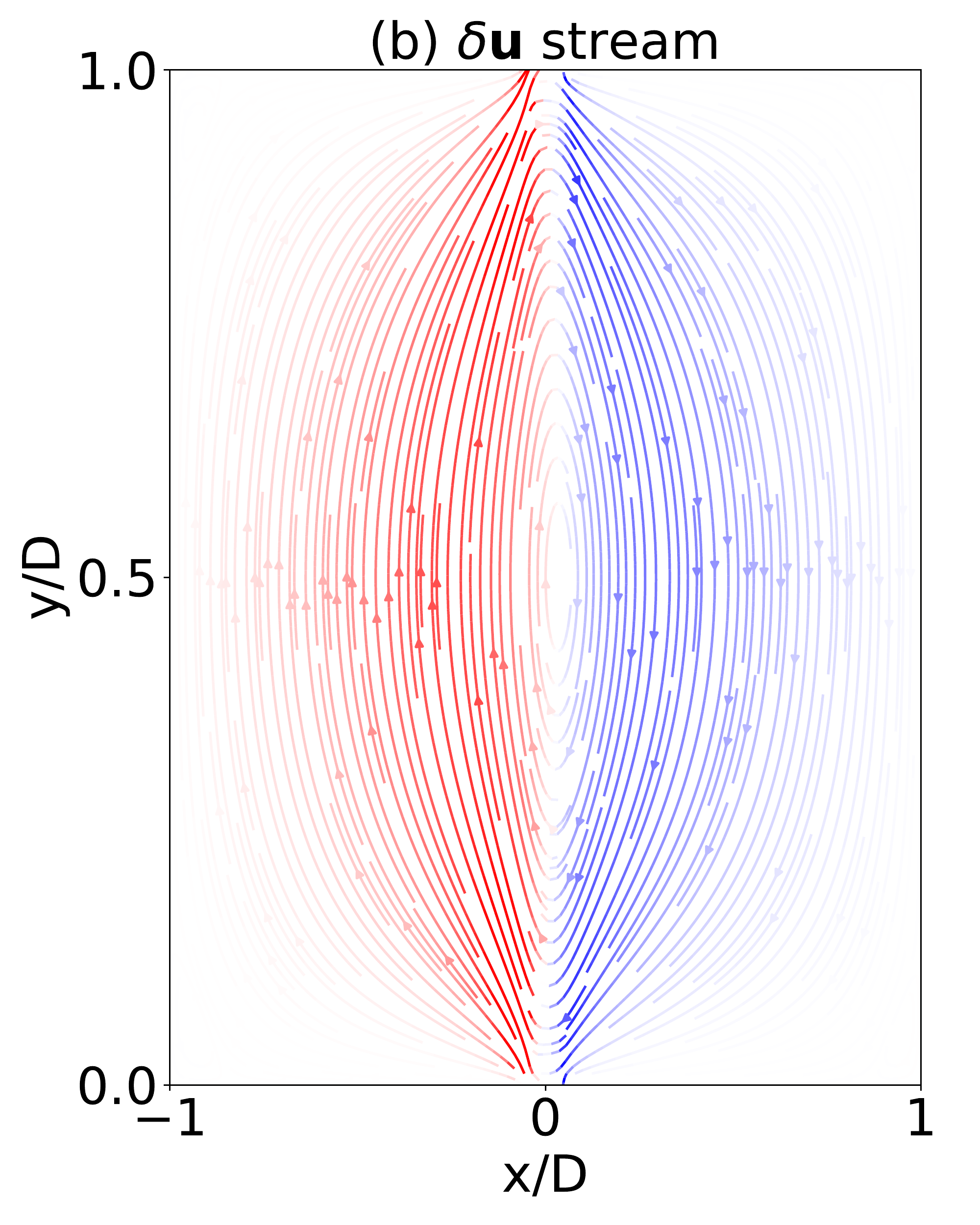}
\caption{Stream plots of (a)  unperturbed $\vec u^\zero$ and (b) $\delta \vec u = \vec u - \vec u^\zero$ . We set $\mathcal{K} = 0.01$}
\label{FigUNum}
\end{figure}

\begin{figure}
\centering
\includegraphics[trim=0cm 0cm 0cm 0cm, clip,width=0.23\textwidth]{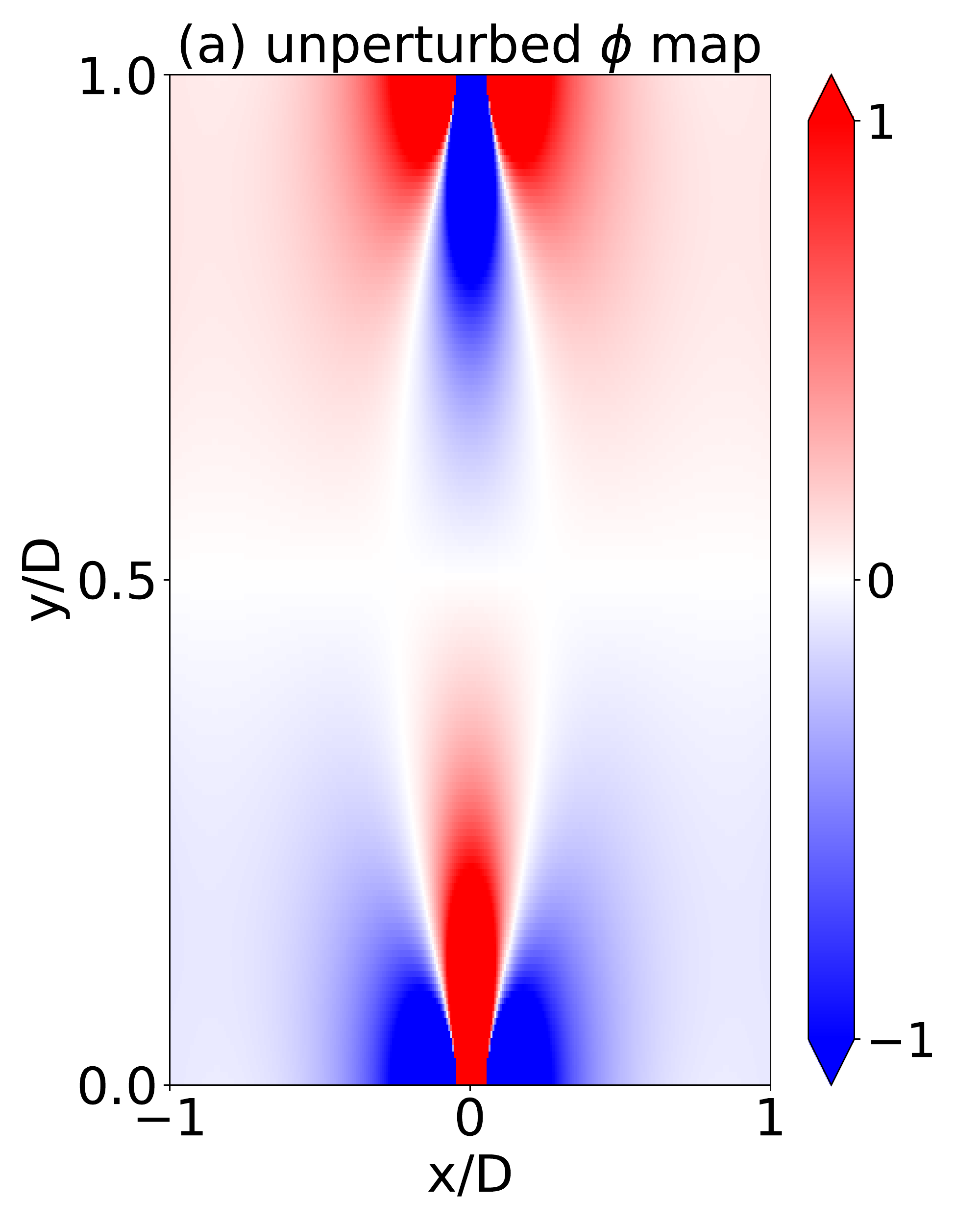}
\includegraphics[trim=0cm 0cm 0cm 0cm, clip,width=0.23\textwidth]{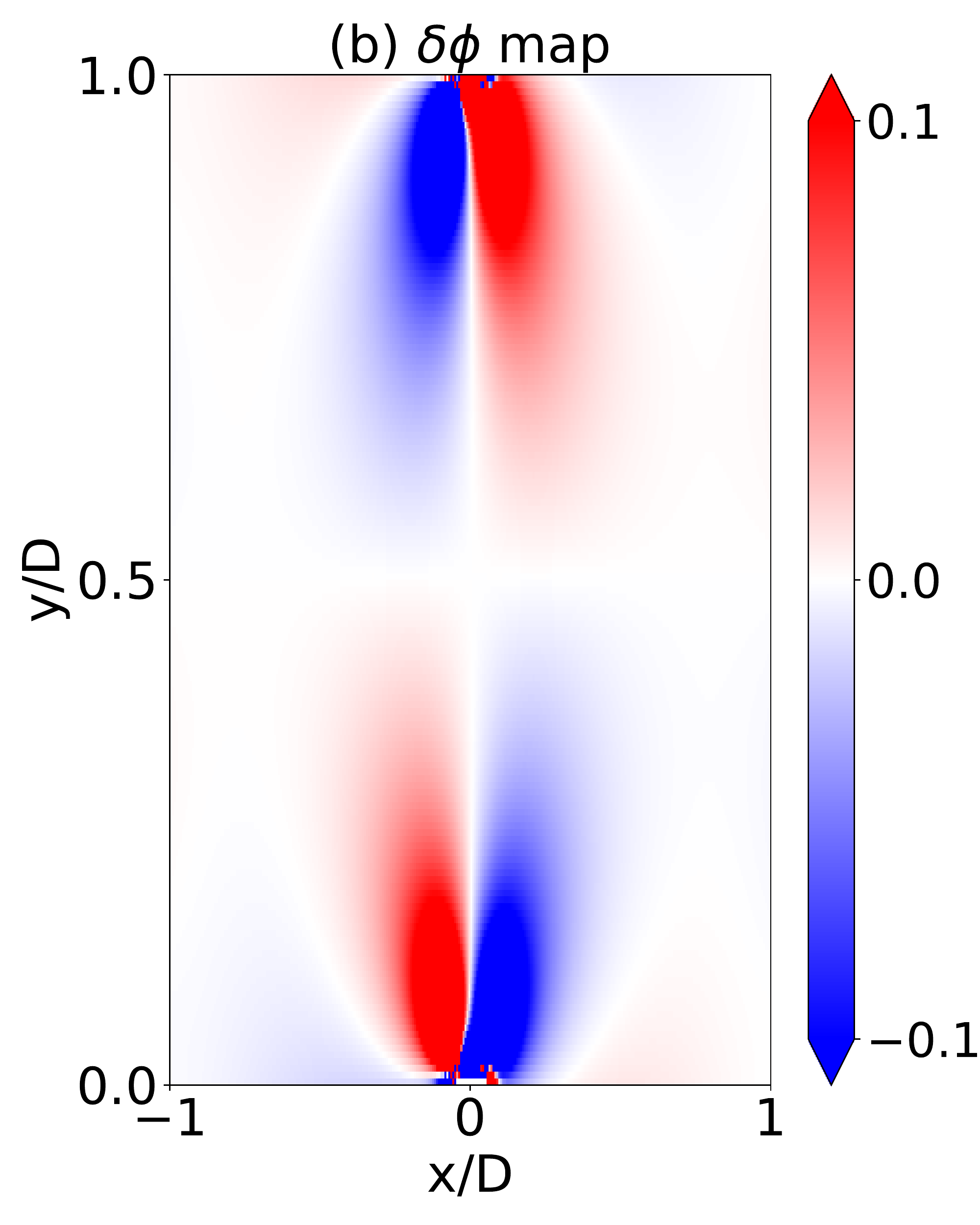}
\caption{Potential profiles (a) $\phi^{(0)}$ without the Berry curvature and (b) $\delta \phi = \phi - \phi^\zero$. We set $\mathcal{K} = 0.01$}
\label{FigPhiNum}
\end{figure}

In Fig.~\ref{Figphi}(a), we show the unperturbed profile of $\phi^\zero$ reproducing previous works~\cite{Pellegrino2016,Falkovich2017}. It is
well known that the sign change of the potential $\phi^\zero$ as a function of $x$
is a hallmark of viscous flow that leads to a negative non-local
resistance even without current backflow. On the other hand, $\delta
\phi$ is anti-symmetric in $x$ and changes sign multiple times along the
azimuthal direction [see Fig.~\ref{Figphi}(b)]. The asymmetric profile of $\delta \phi$ reflects an electric dipole moment induced by the Berry curvature as shown in  the $\nabla^2\delta \phi$ plot in the inset of Fig.~\ref{Figphi}(b).

Both $\phi^\zero$ and $\delta\phi$ formally diverge at the origin as a result of imposing a current profile of the form $u_y=\delta(x)$ at $y=0$. This singularity
and the anti-symmetry of $\delta\phi$ greatly deform the total potential $\phi$ near the
origin even for small $\mathcal{K}=0.1$. Meanwhile, further away from
the origin, $\phi$ resembles the unperturbed potential $\phi^{\zero}$ because $\delta\phi$
decays faster than $\phi^\zero$ [see Eq.~\eqref{eq:deltasol}]. Scanning the potential along $x$ close to
the origin at $y=0.03 D$, we can compare the potential profile with and
without Berry curvature in Fig.~\ref{Fignonlocal}(a). The potential
with Berry curvature $\mathcal{K}=0.1$ (red) reaches significantly larger
values than the one without Berry curvature (blue). Importantly, near $x=0$
its sign changes, in contrast to the symmetric profile of
$\phi^\zero$. From this, we can deduce the non-local resistance defined as
$R_{\rm nl}=\lb\phi(x)-\phi(0)\rb/I$ in
Fig.~\ref{Fignonlocal}(b). Without Berry curvature, $R_{\rm nl}$ only shows negative values when sweeping along the $x$ axis. For nonzero Berry curvature, in contrast, we observe areas of both positive and negative $R_{\rm nl}$ when crossing $x=0$. While the results of our perturbative analysis only hold for a $\delta$-shaped inlet current, we expect that these qualitative features of $\phi$ and $R_{\rm nl}$ will not change even if we consider a finite inlet width.

\section{Numerical solution of the NSE}
To check the validity of the perturbative results, we solve Eq.~\eqref{eq:NSnodim} using the finite-element method. We first rewrite Eq.~\eqref{eq:NSnodim} in a variational form by multiplying the equation by a vector-valued test function $\mathbf{v}$ and then integrating over the entire system volume $\Omega$. Finally, the $\nabla^2\vec u$ term is integrated by parts using Green's identity, and we obtain the variational form
\begin{align}
-\mathcal{K}\integral{\Omega}{}{d\mathbf{r}}\ \vec v\cdot\left(\nabla\phi\times\boldsymbol{\mathcal{B}}\cdot\nabla\right)\mathbf{u} + \integral{\Omega}{}{d\mathbf{r}}\ \vec v\cdot\lp \nabla\phi-\vec u\rp  & \nonumber \\ +\integral{\Omega}{}{d\mathbf{r}}\nabla\mathbf{u}\cdot \nabla \mathbf{v}  - \integral{\partial\Omega}{}{ds}\mathbf{v}\cdot\frac{\partial\mathbf{u}}{\partial\hat{\mathbf{n}}}
&= 0,
\label{Eq:VariationalFormulation1}
\end{align}
where  $\hat{\mathbf{n}}$ is the unit normal pointing out from the surface. Similarly we write the continuity equation in a variational form as well,
\begin{equation}
\integral{\Omega}{}{d\mathbf{r}}\left(\nabla\cdot\mathbf{u}\right)q = 0,
\label{Eq:VariationalFormulation2}
\end{equation}
where $q$ is a scalar test function. The variational problem consists in finding $\vec u$ and $\phi$ such that Eqs.~\eqref{Eq:VariationalFormulation1} and \eqref{Eq:VariationalFormulation2} are satisfied for all test functions $\vec v$ and $ q$.

To this end we have to specify the geometry, boundary conditions and
the function spaces. To stay as close a possible to our analytical calculation, we choose a rectangular geometry with a narrow inlet and outlet at the lower and upper edge, respectively~\cite{Levitov2016, Pellegrino2016}. Thus
our geometry is a rectangle of width $W = 2 D$ and height $H = 1 D$,
where $D$ is the diffusion length. The inlet and
outlet have widths $w_{\rm in/out} = 0.1D $. The boundary conditions on the inlet are $u_y^{\rm in} =
 1$ and $\phi_{\rm in} = 1$ while on the outlet we choose $u_y^{\rm out} = 1$ (as required by the continuity equation in an incompressible medium) and
$\phi_{\rm out} = -1$. On the rest of the boundary we apply no-slip
boundary conditions. For the finite-element analysis, we use the space of Lagrange polynomials of second order as the function space for the velocity and the space of Lagrange polynomials of first order as the function space for the potential. Together these form a Taylor-Hood element appropriate for the numerical solution of Navier-Stokes equations. We have numerically implemented the problem using the \textit{FEniCS} package \cite{LangtangenLogg2017}.

We present the results of the numerical calculations in
Figs.~\ref{FigUNum} and~\ref{FigPhiNum}. Figure~\ref{FigUNum}(a)
shows the velocity profile $\vec u^\zero$ for a system without Berry
curvature. This can be directly compared to the
results of Refs.~\onlinecite{Levitov2016,Pellegrino2016}. The flow exhibits
a pair of whirlpools due to the backflow caused by the boundaries and the location of the whirlpools is comparable with what has been obtained in Ref.~\onlinecite{Levitov2016}.

Introducing a small Berry curvature ($\mathcal{K}=0.01$), we obtain small change in $\vec u$. This is shown in Fig.~\ref{FigUNum}(b), where we have plotted the difference $\delta \vec u=\vec u-\vec u^\zero$. Most notably, $\delta \vec u$ shows a circular flow with a whirlpool in
the center. The upward velocity in the right is reduced while in the
left it is increased by the Berry curvature. As a result, the whirlpools will shift towards the left of the geometry.

We note that we did not observe an additional whirlpool at the inlet, in contrast to what the perturbative solution in the half-plane
geometry predicted. Generally speaking, a quantitative agreement between analytical and numerical results is not expected because of the use of different geometries (infinite versus finite systems, infinitesimal vs.~finite inlet and outlet widths), as well as the small $\mathcal{K}$ value chosen for the numerical simulation. Nevertheless the different
symmetries of $\vec u$ and $\delta \vec u $ in $x$ are an excellent agreement with the analytical solution for the
half-plane geometry, resulting in an asymmetric flow profile $\vec u$. We note that
large values for $\mathcal{K}$ make Eq.~\eqref{Eq:VariationalFormulation1}
highly nonlinear, and the ensuing onset of turbulence makes it challenging to reach convergence in the numerical solution.

The profile of the electric potential in the rectangular geometry qualitatively reproduces previous
results~\cite{Levitov2016,Pellegrino2016} as shown in
Fig.~\ref{FigPhiNum}(a). While the analytical solution rests on assuming a $\delta$-shaped current inflow, the inlet and outlet in our numerical solutions have finite widths. As a consequence, there is no singularity of the potential at the inlet and
outlet. In contrast, sign changes of the
potential near the inlet and outlet agree with the analytical predictions and can be seen as a signature of viscous flow.

The Berry curvature modifies the potential profile as shown in Fig.~\ref{FigPhiNum}(b). Similar to previous analysis, the change of potential $\delta\phi =\phi-\phi^\zero$, plotted in Fig.~\ref{FigPhiNum}(b) makes it easier to recognize the effect of the Berry curvature. Focusing on the lower half, the asymmetric profile
of $\delta \phi$ will expand the blue (red) region of $\phi^\zero$ in
the right (left) side. As a result, the tails of the positive potential (red) near the bottom and of the negative potential (blue) near the top will be diverted to the left. The symmetries of $\phi^\zero$ and $\delta\phi$ in the rectangular geometry are consistent with those in half-plane geometry obtained from perturbation theory.

\section{Conclusions}

In summary, we have derived the Navier-Stokes equation for a two-dimensional electron liquid in the
presence of Berry curvature in a system with broken time-reversal
symmetry. For a Fermi energy slightly above the gap and at an
intermediate temperatures, the longitudinal viscous current can exceed the Hall
current  at small electric fields. This allows for the observation of
an unconventional one-dimensional Poiseuille flow in which the maximum velocity deviates
from the center of the channel due to the Berry curvature. In the case of a two-dimensional geometry, the Berry curvature induces an electric dipole momentum leading to several interesting consequences: current whirlpools as well as asymmetric velocity and potential profiles. The changes of the flow velocity profile and the potential due to the Berry curvature have an opposite symmetry as the corresponding unperturbed quantities. Our analytical results based on a perturbative
method to solve the Navier-Stokes equation have been qualitatively confirmed by a numerical study based on
finite-element methods. We have shown that the presence of Berry curvature can be analyzed experimentally by a non-local resistance measurement in the vicinity of the inlet current.

\begin{acknowledgments}
The authors acknowledge helpful discussions with K.~Moors. All authors  acknowledge support by the National Research Fund, Luxembourg under grants ATTRACT 7556175, CORE 13579612, and CORE 11352881.
\end{acknowledgments}

\bibliography{hydro2}

\appendix

\section {Solution of homogeneous Navier-Stokes equation}
Here we show the procedures used to solve the homogeneous Navier-Stokes equation following Ref.~\onlinecite{Pellegrino2016}. We start with the dimensionless Eq.~\eqref{eq:zero} or the homogeneous part of Eq.~\eqref{eq:delta}
\be
-\nabla\phi^\zero+ \nabla^2\vec u^\zero-\vec u^\zero= 0. \label{eq:azero}
\ee
The incompressibility condition reads $\nabla\cdot\vec u =0$. Writing out the Fourier components and noting that $\partial_x =ik$, we get
\be
-
    \begin{pmatrix}
     (ik) \phi_k^\zero\\
     \partial_y \phi_k^\zero
   \end{pmatrix}
   + (-k^2+\partial_y^2)
   \begin{pmatrix}
     u_{k,x}^\zero\\
     u_{k,y}^\zero
   \end{pmatrix}
   -
   \begin{pmatrix}
     u_{k,x}^\zero\\
     u_{k,y}^\zero
   \end{pmatrix}=0\label{eq:NS-comp}
     \ee
     From Eqs.~\eqref{eq:NS-comp}, we can write a matrix equation:
     \be
     \partial_y
     \begin{pmatrix}
       ku_{k,x}^\zero\\
       ku_{k,y}^\zero\\
       \partial_y u_{k,x}^\zero\\
       k^2 \phi^\zero
     \end{pmatrix}
     =
     k
     \begin{pmatrix}
       0 & 0 & 1 & 0\\
       -i & 0 & 0 & 0\\
      \frac{q^2}{k^2}  & 0 &0 & \frac{i}{k^2}\\
       0& -q^2 & -ik^2 &0
     \end{pmatrix}
     \begin{pmatrix}
       ku_{k,x}^\zero\\
       ku_{k,y}^\zero\\
       \partial_y u_{k,x}^\zero\\
       k^2 \phi^\zero_k
     \end{pmatrix},\label{eq:mat}
     \ee
where $q=\sqrt{1+k^2} $.
     The eigenvalues  are:
     \be
     \lambda_{1,2}=\mp 1,\quad \lambda_{3,4} =\frac{\mp q}{|k|},
     \ee
     and the eigenvectors: $\omega_1=\lp -i,1,i,1\rp^T $, $ \omega_2= \lp -i,-1,-i,1\rp^T$, $\omega_3=\lp -|k|/q,- ik^2/q^2 ,1,0 \rp^T$, and $\omega_4= \lp |k|/q,- ik^2/q^2,1,0 \rp^T$.

     The solutions of Eq.~\eqref{eq:mat} are linear combinations of the four eigenstates,
     \be
     \begin{pmatrix}
       ku_{k,x}^\zero\\
       ku_{k,y}^\zero\\
       \partial_y u_{k,x}^\zero\\
       k^2 \phi_k^\zero
     \end{pmatrix}
     =\sum_{i=1}^4 a_i \omega_i e^{k\lambda_i y}\label{eq:solx}
     \ee
     where $a_i$ is the coefficient satisfying the boundary conditions.
  For the unperturbed solution, the injected current as shown in Fig.~\ref{Fig0} becomes one of the (dimensionless) boundary conditions in real space,
     \be
     u_y^\zero(x=0,y=0)= \delta(x).\label{eq:bc1x}
     \ee
     In Fourier space it becomes,
      \be
     u_{k,y}^\zero(y=0)= 1.\label{eq:bc1-k}
     \ee
     The general form of the boundary condition at the edge $y=0$ reads
     \be
     \lp \partial_yu_x^\zero +\partial_x u_y^\zero \rp \bigg|_{y=0}= \frac{1}{l_b} u_x^\zero (y=0),\label{eq:bc2x}
     \ee
     where the limit $l_b=0$ corresponds to no-slip boundary conditions, while $l_b\to \infty$ describes no-stress boundary conditions.
     In the Fourier space, these becomes
     \be
     \lp \partial_yu_x^\zero +(ik) \rp \bigg|_{y=0}= \frac{1}{l_b} u_x^\zero (y=0).\label{eq:bc2-k}
     \ee
     We also require $u_x^\zero$ and $u_y^\zero$ to vanish at infinity,
     \be
     u_x^\zero(y=\infty)=0,\quad u_y(y=\infty)=0. \label{eq:bc3x}
     \ee
     Because of Eq.~\eqref{eq:bc3x}, we need to choose different signs of $k$ such that each factor $e^{\lambda_ik y}$ decays to zero at $y\to\infty$. Equation~\eqref{eq:solx} then reads,
     \be
     \begin{pmatrix}
       ku_{k,x}^\zero\\
       ku_{k,y}^\zero\\
       \partial_y u_{k,x}^\zero\\
      k^2 \phi^\zero
     \end{pmatrix}
     =
     \left\{
     \begin{matrix}
       a_1 \omega_1 \exp(-ky) +a_3 \omega_3 \exp\lp-q y\rp & \quad k\ge 0\\
       a_2 \omega_2 \exp (ky)+ a_4 \omega_4 \exp\lp-q y\rp & \quad k<0
     \end{matrix}
     \right..
     \label{eq:sol2}
     \ee
      For $k>0$, we use Eq.~\eqref{eq:bc1-k} and \eqref{eq:bc2-k} to get,
     \bea
     a_1 &=& -k\lb\frac{k l_b (k^2+q^2)+q|k|}{k(k+k^2 l_b -l_b q^2)-q|k|}\rb\\
     a_3 &=& ikq^2\lb \frac{(1+2 k l_b)}{k(k+k^2 l_b-l_b q^2)-q|k|}\rb
     \eea
     Similarly, we obtain for $k<0$,
\bea
     a_2 &=& k\lb\frac{k l_b (k^2+q^2)-q|k|}{k(-k+k^2 l_b -l_b q^2)+q|k|}\rb\\
     a_4 &=& ikq^2\lb \frac{(-1+2 k l_b)}{k(-k+k^2 l_b-l_b q^2)+q|k|}\rb
     \eea

     We write the unperturbed solutions as follows,
     \begin{align}
     u_{k,x}^\zero &= \frac{ik}{|k|} \lb\frac{[l_b(k^2+q^2)+q]e^{-|k| y}}{|k|-q + l_b(k^2-q^2)} - \frac{[q(1+2|k|l_b)]e^{-qy}}{|k|-q + l_b(k^2-q^2)}\rb\nn
     u_{k,y}^\zero  &=  \lb\frac{-[ l_b (k^2+q^2) +q]e^{-|k|y}}{|k|-q + l_b(k^2-q^2)} + \frac{[|k|(1+2|k|l_b)]e^{-qy}}{|k|-q + l_b(k^2-q^2)}\rb \nn
     \phi_k^\zero &=-\frac{1}{|k|}\lb\frac{ l_b (k^2+q^2) +q}{|k|-q + l_b(k^2-q^2)}e^{-|k| y}\rb
     \end{align}
    For no-slip boundary conditions ($l_b\to 0$), we then obtain Eq.~\eqref{eq:zerosol},
     \bea
     u_{k,x}^{(0)}&=& \frac{i kq}{|k|(|k|-q)}\lb  e^{-|k|y}- e^{-qy}\rb\nn
     u_{k,y}^{(0)}&=&-\frac{1}{(|k|-q)} \lb q e^{-|k|y}- |k| e^{-qy} \rb\\
     \phi_k^{(0)} &=& -\frac{1}{|k|} \frac{q}{|k|-q} e^{-|k|y}  \nonumber
     \eea
     For the homogeneous solution of the first-order correction Eq.~\eqref{eq:delta}, we use the same procedure up to Eq.~\eqref{eq:sol2} and then use boundary conditions $\delta \vec u=0$ at $y=0$.

    \section{Numerical evaluation of $\mathcal{B}_z$}
     Here we show the numerical evaluation of $\mathcal{B}_z$ in a gapped Dirac system,
     \bea
     \mathcal{B}_z&=& \frac{e}{\hbar}\sum_p p_x^2 \Omega_z \lp\frac{-\partial f^0}{\partial \eps}\rp, \quad \eps =\sqrt{v^2p^2+\Delta^2}\nn
     &=& \frac{e}{\hbar}\frac{1}{(2\pi\hbar)^2}\int d^2\vec p p^2 \cos^2 \theta \frac{v^2\hbar^2\Delta}{2\eps^3} \lp\frac{e^{(\eps-\mu)/k_BT}}{k_BT(1+e^{(\eps-\mu)/k_BT})^2}\rp,\nn
     &=& \frac{e}{h}\frac{\Delta^2}{v^2k_BT}b(\tilde\mu, \tilde T),\label{eq:bz}
     \eea
     where
     \be
     b(\tilde\mu, \tilde T)=\frac{1}{4}\int_1^{\infty} dx \lp 1-\frac{1}{x}\rp \frac{e^{(x-\tilde\mu)/\tilde T}}{(1+e^{(x-\tilde\mu)/\tilde T})^2},
     \ee
     and $\tilde \mu=\mu/\Delta$ and $\tilde T=k_BT/\Delta$. In Eq.~\eqref{eq:bz} we have transformed the $\vec p$-integral over momenta into an integral over energies $\eps$.
\end{document}